\documentclass[a4paper,11pt]{article}
\usepackage{jheppub}
\usepackage{lineno}
\usepackage{amsmath,amssymb,slashed,makecell}

\author[a,b]{Maksym Ovchynnikov}
\affiliation[a]{Institut für Astroteilchen Physik, Karlsruher Institut für Technologie (KIT), Hermann-von-Helmholtz-Platz 1, 76344 Eggenstein-Leopoldshafen, Germany
}

\affiliation[b]{Instituut-Lorentz for Theoretical Physics, Universiteit Leiden, Niels Bohrweg 2, 2333 CA Leiden, The Netherlands}

\author[a,c]{Jing-Yu Zhu}

\affiliation[c]{School of Physics and Astronomy and Tsung-Dao Lee Institute, Shanghai Jiao Tong University, 800 Dongchuan Rd, Shanghai 200240, China}
\emailAdd{maksym.ovchynnikov@kit.edu,zhujingyu@sjtu.edu.cn}

\begin{document}

\title{Search for the dipole portal of heavy neutral leptons at future colliders}

\abstract
{In this paper, we study the potential of future colliders to explore the parameter space of heavy neutral leptons (HNLs) through the dipole portal. We consider hadron colliders such as the LHC in the high luminosity phase and FCC-hh, and lepton colliders, such as FCC-ee. We consider various signatures for the HNLs, including the missing energy signature and displaced decays, and discuss the complementarity between the hadron and lepton colliders. We propose new selection rules which may significantly reduce the background events in FCC-ee. In particular, we find that thanks to a much clearer environment, FCC-ee may search for the HNLs with masses up to $\simeq 30\text{ GeV}$ and proper lifetimes $c\tau_{N}\gtrsim 1\text{ cm}$, which is well beyond the reach of the experiments to be launched in the next decade.}

\maketitle
\flushbottom

\section{Introduction}
\label{sec:introduction}

The neutrino dipole portal of the heavy neutral lepton (HNL) originates from a six-dimension operator~\cite{Magill:2018jla}:
\begin{equation}
    \mathcal{L}_{\text{dipole}} = \bar{L}(d_{\mathcal{W}}\mathcal{W}_{\mu\nu}^{a}\tau_{a}+d_{B}B_{\mu\nu})\tilde{H}\sigma^{\mu\nu}N+\text{h.c.} \;,
    \label{eq:operator}
\end{equation}
where $L$ denotes the SM lepton doublet, ${\cal W}_{\mu\nu}^a$ and $B_{\mu\nu}$ stand for the $SU(2)_{L}$ field strength tensor, $\tau_a = \sigma_a/2$ with $\sigma_a$ being Pauli matrix, $\tilde H = i \sigma_2 H^{*}$ is the conjugate Higgs field, and $N$ is the HNL gauge singlet.\footnote{If having two HNLs $N_{1},N_{2}$, it may be possible to consider dimension-5 operators that couple $N_{1},N_{2}$ to the gauge field strength (see, e.g.~\cite{Barducci:2022gdv}).}
After the spontaneous symmetry breaking, we have
\begin{equation}
\begin{split}
\mathcal{L}_{\text{dipole}}^{\text{eff}}= 
 \bar{\nu}_{\alpha L}(d_{\alpha}F_{\mu\nu}-d_{Z}Z_{\mu\nu})\sigma^{\mu\nu}N +d_{W}\bar{l}_{L}W_{\mu\nu}\sigma^{\mu\nu}N + \text{h.c.},
\end{split}   
\label{eq:dipole-portal-couplings}
\end{equation}
where 
\begin{equation}
d_{\alpha}= \frac{v \cos\theta_{W}}{\sqrt{2}} (d_B+ \frac{\tan\theta_{W}}{2} d_{\cal{W}}),~ d_Z=\frac{v \cos\theta_{W}}{\sqrt{2}} (\frac{d_{\cal W}}{2}- \tan\theta_{W} d_B), ~d_W=\frac{v}{2} d_{\cal{W}},
\label{eq:couplings}
\end{equation}
with $v$ being the Higgs vacuum expectation value,
the couplings $d_{Z},d_{W}$ vary within the range
\begin{equation}
|d_{Z}/d_{\alpha}| \in (0, \cot\theta_{W}), \quad |d_{W}/d_{\alpha}|\in \left(0,\frac{\sqrt{2}}{\sin\theta_{W}}\right) 
\label{eq:dWtodgamma-max}
\end{equation}
and $d_{\alpha}$ stands for the dipole coupling with $\alpha = e,\mu,\tau$\footnote{Note that it is illustrated in Ref. \cite{Brdar:2020quo} how to construct UV-complete models which give rise to the operators in Eq.~\eqref{eq:dipole-portal-couplings} at high energy scale $\Lambda$. The TeV-scale leptoquark model can make good compromises between large values of $d_\alpha$ and small values of active neutrino masses, where the phenomenology of HNL is dominated by the dipole portal instead of the mixing one. In Eq.~\eqref{eq:operator}, $d_{\cal{W}}$ and $d_B$ can be rewritten as $C_{\cal{W}}/\Lambda^2$ and $C_{B}/\Lambda^2$, respectively, where $C_{\cal{W}}$ and $C_{B}$ are Wilson coefficients dependent on lepton flavor. It is easy to translate the constraint of $d_{\alpha}$ to that of $C_{\cal{W}}$ and $C_{B}$ from Eq.~\eqref{eq:couplings}. Taking any fixed values of $|d_Z/d_{\alpha}|$ and $|d_W/d_{\alpha}|$ within the two ranges of Eq.\eqref{eq:dWtodgamma-max} directly leads us to $C_{W,B}/\Lambda^2$ proportional to $d_{\alpha}$.}. 

The first works discussing this model were motivated by the MiniBooNE and LSND anomalies~\cite{Gninenko:2009ks,Gninenko:2010pr}. 
Since then, many studies have been made on the constraints and sensitivities of future experiments to the dipole portal, especially very recently~\cite{Shoemaker:2020kji,
Plestid:2020vqf,Jodlowski:2020vhr,Atkinson:2021rnp,Schwetz:2020xra,
Dasgupta:2021fpn,Ismail:2021dyp,Miranda:2021kre,Bolton:2021pey, Arguelles:2021dqn,
Mathur:2021trm,Li:2022bqr,Zhang:2022spf,Huang:2022pce,Gustafson:2022rsz,
Kamp:2022bpt,Abdullahi:2022cdw,Delgado:2022fea,Ovchynnikov:2022rqj,Abdullahi:2022jlv,Zhang:2023nxy}.

\begin{figure}[!ht]
    \centering
    \includegraphics[width=0.7\textwidth]{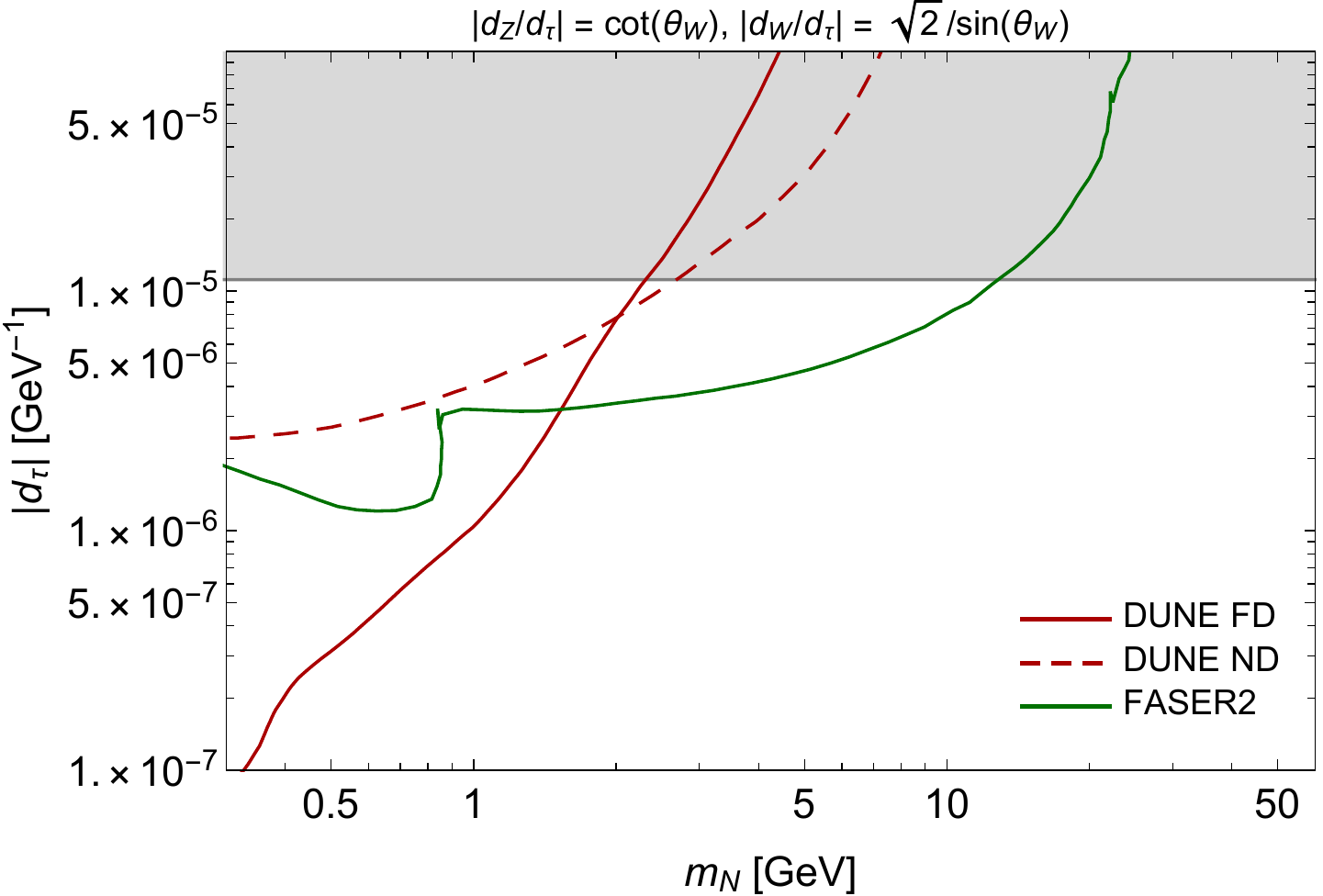}
    \caption{Sensitivity of neutrino factories (such as DUNE~\cite{Ovchynnikov:2022rqj} and FASER2~\cite{Jodlowski:2020vhr}) to HNLs with the dipole coupling to $\tau$ neutrinos. The bounds from past experiments (shown as gray region) have been obtained from~\cite{Schwetz:2020xra}.}
    \label{fig:sensitivity-dipole-neutrino-factories}
\end{figure}

The constraints on the HNLs can be briefly summarized as follows~\cite{Schwetz:2020xra,Magill:2018jla}:
\begin{itemize}
\item[--] From the viewpoint of outer space, HNL may influence the stellar cooling process as well as the expansion and cooling of the universe. The latter can be scrutinized by analyzing Big Bang Nucleosynthesis and CMB constraints. The above processes are mainly sensitive to $m_N$ with masses below 
$300$ MeV~\cite{Brdar:2020quo}.
\item[--]  One of the efficient production mechanisms of HNLs is the neutrinos up-scattering off electrons or nucleons\cite{Magill:2018jla,Schwetz:2020xra}.
Therefore, the dipole portal may be constrained in neutrino or dark matter experiments. Examples include CHARM-II, DONUT, NOMAD, LSND, MiniBooNE, SBND, MicroBooNE, CHARM-II, DONUT, NOMAD, Borexino, Super-Kamiokande, IceCube, SHiP, DUNE, COHERENT, NUCLEUS, Xenon1T, SuperCDMS, etc., see~\cite{Magill:2018jla, Schwetz:2020xra, Bolton:2021pey}. Because of the limited energy of neutrino sources, these constraints or sensitivities to HNL masses can be at most at a few GeV.
\item[--] At the energy frontier, it is feasible to probe the existence of much heavier HNL through the dipole portal. A discussion of the constraints coming from the LEP and the LHC has been made in~\cite{Magill:2018jla}.
\end{itemize}
The currently unexplored parameter space may be probed in various future experiments. Relatively light HNLs with masses $m_{N}\lesssim \mathcal{O}(1\text{ GeV})$ may be produced by decays of light mesons such as $\pi,K$, and in neutrino up-scatterings~\cite{Magill:2018jla,Schwetz:2020xra} at neutrino factories. Examples are currently running SND@LHC~\cite{SNDLHC:2022ihg}, FASER$\nu$~\cite{FASER:2020gpr} as well as their future upgrades -- advSND~\cite{Feng:2022inv} and FASER2/FASER$\nu$2~\cite{Jodlowski:2020vhr}, SHiP~\cite{SHiP:2015vad}, and DUNE~\cite{DUNE:2020lwj}. The plot showing sensitivities of DUNE and FASER2 to HNLs coupled to $\tau$ neutrinos is shown in Fig.~\ref{fig:sensitivity-dipole-neutrino-factories}. Another potential probe of the HNL parameter space may come from Belle II~\cite{Zhang:2022spf}.

However, the sensitivity of neutrino factories quickly diminishes with the increase of the HNL mass $m_{N}$. This is because of the quick decrease of the up-scattering production cross-section with $m_{N}$. Therefore, we need other experiments to explore heavier HNLs with small couplings. Future ultrahigh energy neutrino telescope may contribute to the sensitivities of HNL masses as large as 30 TeV~\cite{Huang:2022pce}. To probe masses above 1 GeV and small couplings, one may consider various signatures with HNLs at colliders: lepton colliders, such as FCC-ee~\cite{FCC:2018byv}, CEPC~\cite{CEPCStudyGroup:2018rmc,CEPCStudyGroup:2018ghi}, the muon collider~\cite{Aime:2022flm,Black:2022cth}, and hadron colliders -- the LHC in its high luminosity phase, FCC-eh~\cite{FCC:2018byv} and FCC-hh~\cite{FCC:2018vvp}. 

The sensitivity of CEPC to the HNLs for the particular signature of a photon and missing energy has been obtained in~\cite{Zhang:2023nxy}. However, to the best of our knowledge, a systematic study of the potential of the colliders to explore the dipole portal is lacking.

In this work, we analyze how different signatures with HNLs may be used to search for them at the lepton and hadron colliders. The paper is organized as follows: in Sec.~\ref{sec:HNLs-at-colliders}, we study the phenomenology of HNLs at colliders, including their production and decay, and consider different signatures used to search for these HNLs. We also discuss how the events originating from the dipole coupling may be distinguished from those originating from the HNL mixing with neutrinos. Subsequently, the sensitivities to the dipole portal at hadron and lepton colliders are discussed in Sec.~\ref{sec:hadronic-colliders} and Sec.~\ref{sec:lepton-colliders}, respectively. Finally, in Sec.~\ref{sec:conclusions}, we make conclusions.

\section{HNLs with dipole coupling at colliders}
\label{sec:HNLs-at-colliders}

In this section, we discuss details on the production and decay of HNLs, and how to search for them at colliders. Below, we marginalize over $d_{Z},d_{W}$ by choosing their maximal possible values (see Eq.~\eqref{eq:dWtodgamma-max}), namely, $d_{Z} = \cot(\theta_{W})d_{\alpha}$ and $d_{W}=\sqrt{2}/\sin(\theta_{W})d_{\alpha}$. We also assume the Dirac nature of HNLs. The analysis for the Majorana HNLs would be similar, except for the twice larger decay length for the fixed coupling and mass (see a discussion in~\cite{Schwetz:2020xra}). 

\subsection{Phenomenology}
\label{sec:hnl-phenomenology}

\begin{table}[ht!]
    \centering
    \begin{tabular}{|c|c|c|c|c|}
      \hline Experiment & $N_{W}$ & $N_{Z}$ \\ \hline
        HL-LHC & $6\cdot 10^{11}$ & $1.5\cdot 10^{11}$ \\ \hline FCC-hh & $1.2\cdot 10^{13}$ & $9\cdot 10^{12}$ \\ \hline FCC-ee & $5\cdot 10^{8}$ & $5\cdot 10^{12}$ \\ \hline  
    \end{tabular}
    \caption{The numbers of bosons used in the simplified estimates for the HNL number of events in Sec.~\ref{sec:HNLs-at-colliders}. They are taken from Refs.~\cite{FCC:2018byv,ATLAS:2016fij} (for HL-LHC and FCC-ee), or obtained using MadGraph tree-level simulation~\cite{Alwall:2014hca} (for FCC-hh).}
    \label{tab:accelerator-parameters}
\end{table}

\paragraph{Production.} At colliders, there are two mechanisms of the production of HNLs with masses $m_{N}\gtrsim 1\text{ GeV}$. The first mechanism is direct production: 
\begin{equation}
l^{+}+l^{-} \to N + \nu
\end{equation}
at the lepton colliders, and 
\begin{equation}
p+p\to N+\nu +X, \quad p + p \to N+l+X
\end{equation}
at the hadron colliders. The processes with neutrinos may go via all the possible couplings $d_{\alpha},d_{Z},d_{W}$ in Eq.~\eqref{eq:dipole-portal-couplings}, while the process 
with the charged lepton is via $d_{W}$. The second production mechanism is via decays of  $W$ and $Z$ bosons, 
\begin{equation}
W\to N+l, \quad Z\to N+\nu,
\end{equation}
controlled by $d_{W}$ and $d_{Z}$ 
coupling, respectively. The amounts of these bosons at the LHC, FCC-hh, and the Z-pole mode of FCC-ee are reported in Table~\ref{tab:accelerator-parameters}. In the low-mass limit $m_{N}\ll m_{W/Z}$, the branching ratios of these decay processes behave as
\begin{equation}
    \text{Br}(W\to N+l) \approx \frac{d_{W}^{2}m_{W}^{3}}{12\pi \Gamma_{W}} \approx 6.5\cdot 10^{3}\left(\frac{d_{W}}{\text{GeV}^{-1}}\right)^{2}\;,
\end{equation}
\begin{equation}
    \text{Br}(Z\to N+\nu) \approx \frac{d_{Z}^{2}m_{Z}^{3}}{12\pi \Gamma_{Z}} \approx 8\cdot 10^{3}\left(\frac{d_{Z}}{\text{GeV}^{-1}}\right)^{2}\;.
    \label{eq:branchings}
\end{equation}
For the HNLs with masses below the $W/Z$ mass, the prompt production is suppressed compared to the decay of the heavy bosons. This in particular means that the muon collider, operating at energies much above the $Z$ boson mass, is not as efficient in probing the parameter space of such HNLs as the electron colliders.

\begin{figure}[!t]
    \centering
    \includegraphics[width=0.7\textwidth]{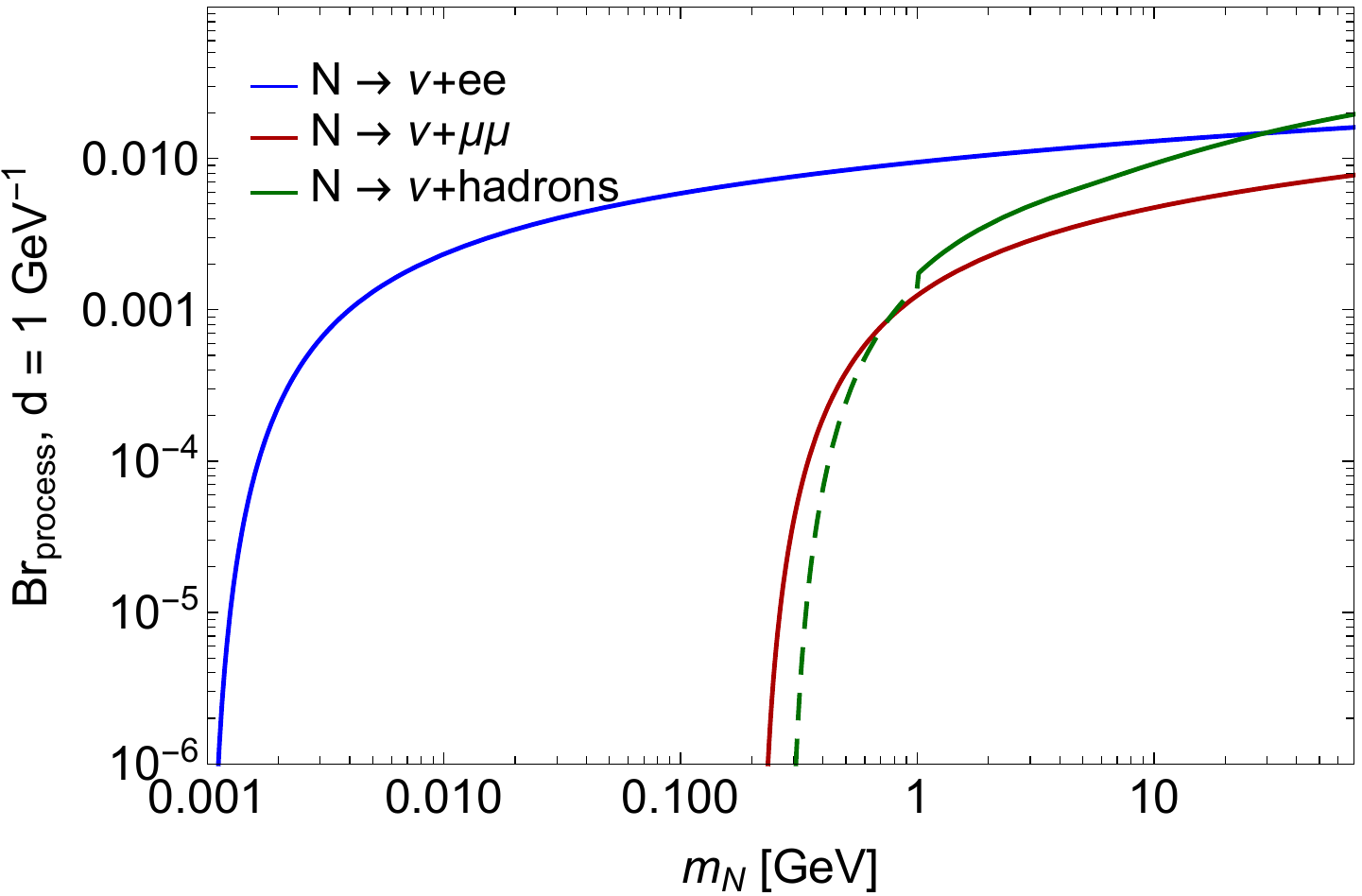}
    \caption{The branching ratios of sub-dominant decays of HNLs: the di-lepton decays $N\to \nu + l^{+} + l^{-}$, where $l = e,\mu$, and the hadronic decays $N\to \nu+\text{hadrons}$, approximated by the decay $N\to \nu+q+\bar{q}$, where $q = u,d,s,c,b$. Note that below $m_{N} \simeq \Lambda_{\text{QCD}}\sim 1\text{ GeV}$, perturbative QCD breaks down, and the corresponding prediction for the hadronic branching ratio becomes invalid.}
    \label{fig:br-ratios}
\end{figure}

\paragraph{Decays.} Decays of HNLs in the mass range $m_{N}\ll m_{W,Z}$ occur mainly via the coupling $d_{\alpha}$. This is because the decay widths for the processes mediated by $d_{W,Z}$ are suppressed by $m_{N}^{4}G_{F}^{2}\ll 1$. The main decay channel is the 2-body decay $N\to \nu+\gamma$. The corresponding decay width is~\cite{Magill:2018jla}
\begin{equation}
    \Gamma_{N\to \nu\gamma} = \frac{
    d_{\alpha}^{2}m_{N}^{3}}{4\pi} \approx \Gamma_{N,\text{tot}}
    \label{eq:decay-width}
\end{equation} 
However, this process does not suit the collider searches that require observing a displaced decay vertex. Instead, sub-dominant decay channels should be considered -- the 3-body decays $N\to f+\bar{f}+\nu$, which occur via the virtual photon. In the limit $m_{N} \gg 2m_{f}$, the corresponding decay width behaves as~\cite{Ovchynnikov:2022rqj}
\begin{equation}
    \Gamma_{N\to \nu f\bar{f}} \approx 
    \frac{\alpha_{\text{EM}}|d_{\alpha}|^{2}m_{N}^{3}Q_{f}^{2}N_{f}}{12\pi^{2}}\left( \log\left[\frac{m_{N}^{2}}{m_{f}^{2}}\right]-3\right),
    \label{eq:dipole-3-body-decay-width}
\end{equation}
where $Q_{f}$ is the electric charge of the fermion $f$, while $N_{f} = 1$ for leptons or $N_{f} = 3$ for quarks. The branching ratios of these processes are shown in Fig.~\ref{fig:br-ratios}.

\subsection{Signatures}
Having discussed the phenomenology of HNLs, we may now consider the signatures at lepton and hadron colliders.

\subsubsection{Hadron colliders}
\label{sec:HNLs-at-hadron-colliders}
\paragraph{Monophoton and missing energy.} One of the possible signatures is the event with a monophoton and missing energy, e.g. 
\begin{equation}
q+\bar{q}\to N + \bar{\nu}, \quad N \to \nu+\gamma, 
\end{equation} 
with the missing energy carried away by neutrinos. This signature has been analyzed in~\cite{Magill:2018jla}, where the authors utilized ATLAS search~\cite{ATLAS:2017nga} performed 
for the dataset corresponding to the integrated luminosity $\mathcal{L} = 36.1\text{ fb}^{-1}$. It was shown that with this type of search, it might be possible to probe the couplings higher than $d\gtrsim 10^{-5}\text{ GeV}^{-1}$. 

Estimating the sensitivity of the LHC in the high luminosity phase is simple. In the accessible parameter space, the HNLs have microscopic proper lifetimes $c\tau_{N}\ll 1\text{ mm}$, and the probability for the HNL to decay is $\approx 1$. Therefore, the number of events scales as 
\begin{equation}
N_{\text{ev}} \sim N_{\text{N,prod}}\times P_{\text{decay}}\times \epsilon_{\text{sel}} \propto \mathcal{L}\times d_{Z}^{2},
\end{equation}
where $N_{\text{N,prod}} \propto \mathcal{L}\times d^{2}_{Z}$ is the total number of the produced HNLs, and $\epsilon_{\text{sel}}$ is the selection efficiency, which we assume will not affect the scaling of the number of events with the HNL model parameters. Given that the background also scales with $\mathcal{L}$, the lower bound of the sensitivity in the plane $m_{N}$--$d$ changes as $\mathcal{L}^{-1/4}$. For HL-LHC, $\mathcal{L}_{\text{HL}} = 3000\text{ fb}^{-1}$, and therefore it should probe a factor of $(3000/38)^{1/4} \approx 3$ smaller couplings than it was derived in~\cite{Magill:2018jla} for the old dataset. We show the corresponding bounds in Figs.~\ref{fig:Nevents-hadronic-colliders} and~\ref{fig:final-sensitivity}.

\paragraph{Displaced vertices.} HNLs with decay lengths $l_{\text{N,decay}}\gtrsim \mathcal{O}(1\text{ mm})$ and masses $m_{N} < m_{W}$ may be searched using displaced vertex (DV) techniques. An example of such a DV scheme is the scheme~\cite{CMS:2022fut} used at CMS to look for HNLs with the mixing coupling. The scheme utilizes the process chain 
\begin{equation}
    p + p \to W+X, \quad W \to N+l, \quad N \to l^{'+} + l^{''-}+\nu,
\end{equation}
where $l,l',l''$ are electrons or muons. To discriminate HNLs from backgrounds, it is required to detect the final state leptons $l',l''$, and the prompt lepton $l$. These particles must have kinematic properties that satisfy some selection criteria. Examples of such properties are large enough transverse momentum and transverse impact parameters. We will discuss the selection in more detail in Sec.~\ref{sec:hadronic-colliders-sensitivity}.

For HNLs with masses $m_{N}<m_{\eta}$, additional production channels by decays of mesons open up. In particular, neutral mesons such as $\pi^{0}/\eta$ may promptly decay into an HNL and a neutrino, $\pi^{0}/\eta\to N+\nu$~\cite{Ovchynnikov:2022rqj}. There are two complications with using these channels. First, although $N_{\pi,\eta}/N_{W/Z}\sim 10^{7}$~\cite{Kling:2021fwx}, the branching ratio of these decays is $\sim 10^{-4}(d_{\alpha}/\text{GeV}^{-1})$ (compare to~\eqref{eq:branchings}, where the branching ratios are parametrically very large); the resulting amounts of the HNLs from heavy bosons and the mesons are thus similar. Second, there is no clear method of distinguishing the events with feebly-interacting particles from light mesons from SM events; in contrast, the events with decays of $W/Z$ may be distinguished by their specific topology and large transverse momenta of the outgoing leptons. Nevertheless, this production channel may be used by far-forward LHC-based experiments such as FASER or FACET, which are located far enough to eliminate the SM background without imposing any requirement on the events with HNLs.

\subsubsection{Lepton colliders}
\label{sec:HNLs-at-lepton-colliders}
\paragraph{Monophoton and missing energy.} Similarly to the hadron colliders, it may be possible to search for the HNLs via the missing energy signature at the lepton colliders. The process of interest is~\cite{Magill:2018jla}
\begin{equation}
l^{+}+l^{-} \to N+ \bar{\nu}, \quad N \to \gamma + \nu
\end{equation}
Similarly to the case of the analogical search at the LHC, it may be possible to derive the sensitivity of FCC-ee from the result of the older searches at DELPHI. The upper bound from LEP on the cross-section of the process $e^{+}e^{-}\to \gamma+\text{inv}$ obtained at the $Z$ pole mode is~\cite{L3:1992cmn,OPAL:1994kgw,DELPHI:1996drf} $\sigma_{\text{mono-}\gamma}^{\text{DELPHI}} = 0.1\text{ pb}$, where for the energy of the photon and its polar angle it was required $E_{\gamma}>0.7\text{ GeV}$ and $|\cos(\theta_{\gamma})|<0.7$. Given the similar background at FCC-ee and LEP (and in particular that both LEP and FCC-ee are free from pileup), and assuming conservatively the same detector properties of FCC-ee as for LEP, the lower bound of the sensitivity of FCC-ee would be 
\begin{equation}
\frac{\sigma_{\text{mono-}\gamma}^{\text{FCC-ee}}}{\sigma_{\text{mono-}\gamma}^{\text{FCC-ee}}} \simeq \left(\frac{\mathcal{L}_{\text{Z-pole}}^{\text{LEP}}}{\mathcal{L}_{\text{Z-pole}}^{\text{FCC-ee}}}\right)^{\frac{1}{4}} = 3.3\cdot 10^{-2}
\end{equation}
where $\mathcal{L}_{\text{Z-pole}}^{\text{LEP}} = 0.2 \text{ fb}^{-1}$ and $\mathcal{L}_{\text{Z-pole}}^{\text{FCC-ee}} = 150\cdot 10^{3} \text{ fb}^{-1}$. 

We show the corresponding sensitivity in Figs.~\ref{fig:lepton-colliders-sensitivity} and~\ref{fig:final-sensitivity}. Note that this simple estimate roughly agrees with the sensitivity of CEPC recently computed in~\cite{Zhang:2023nxy}.

\begin{figure}[!h]
    \centering
\includegraphics[width=0.7\textwidth]{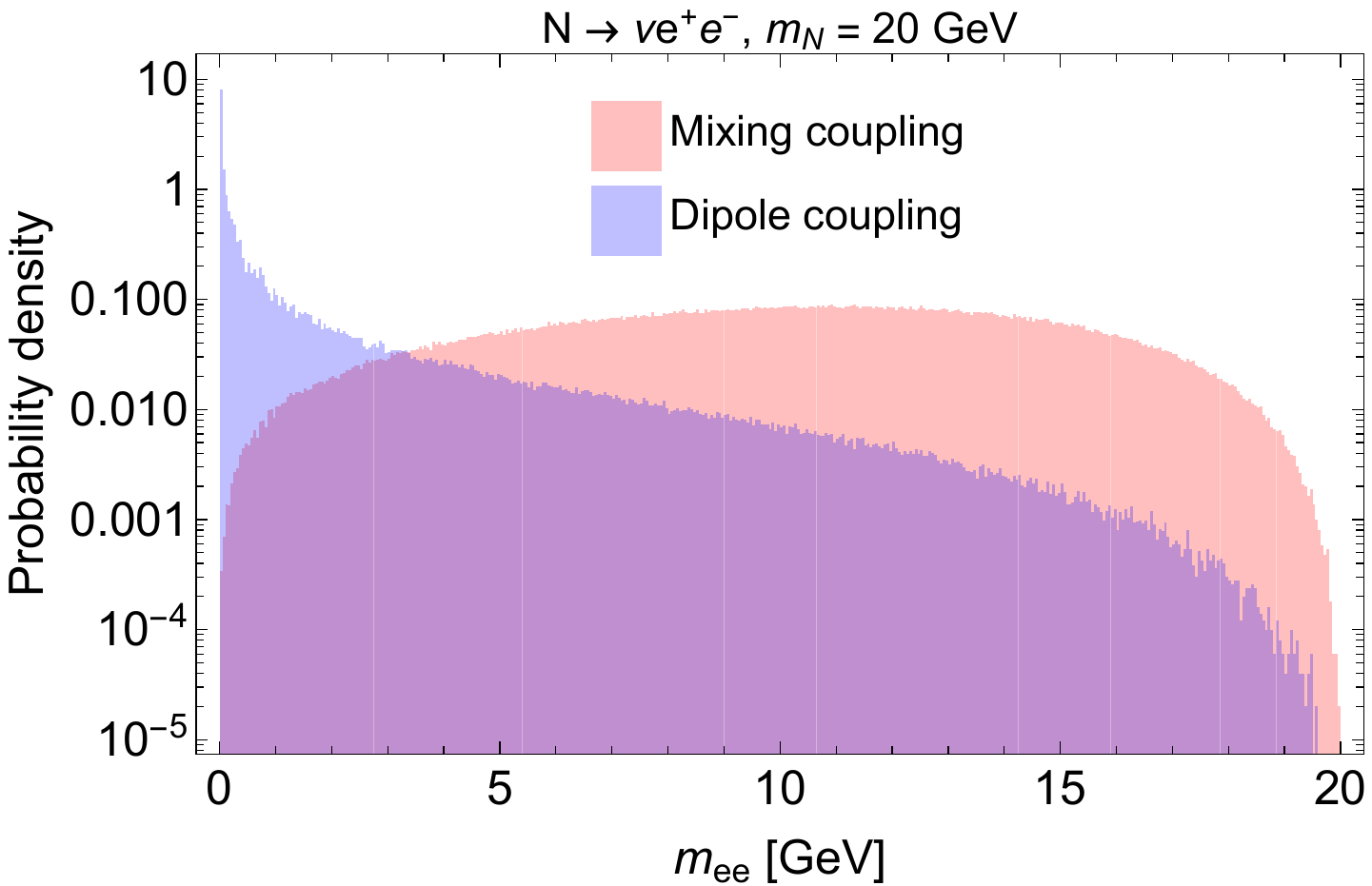}
    \caption{The distribution of the electron-positron pair from the HNL decay $N\to \nu e^{+}e^{-}$ in the invariant mass $m_{ee} = \sqrt{(p_{e^{+}}+p_{e^{-}})^{2}}$, assuming the mixing (the red histogram) or the dipole (the blue histogram) coupling of the HNL to the SM particles.}
    \label{fig:dipole-vs-mixing-distinguishing}
\end{figure}

Let us now discuss how to distinguish leptonic decays of the HNLs with mixing or the dipole couplings. The simplest way would be to check the presence of the leptons of different flavors in the lepton pair: such type of decays is common for the HNLs with the mixing coupling (it occurs via the charged current)~\cite{Bondarenko:2018ptm, Ding:2019tqq,Shen:2022ffi} but is highly suppressed for the HNLs with the dipole coupling. Another way would be to compare the distribution of the lepton pair in invariant mass. For the dipole coupling, the leptons appear via a virtual photon. Therefore, the distribution has the maximum at $m_{\text{inv}} = 2m_{e}$ and quickly drops with the increase of $m_{\text{inv}}$. In contrast, for the mixing coupling, the mediator is a heavy $W/Z$, the corresponding propagator is a constant, and the distribution is rather flat in the range $0<m_{\text{inv}}<m_{N}$, see Fig.~\ref{fig:dipole-vs-mixing-distinguishing}.

\paragraph{Displaced decays}
Another way to search for HNLs may be to look for the displaced decays. Unlike the LHC, lepton colliders have a much cleaner background; in particular, no pile-up events~\cite{An:2018dwb,Boscolo:2019awb}. Therefore, it may be much simpler to distinguish a hypothetical SM background and the signal from decaying HNLs. In particular, instead of searching for the events with prompt leptons, one may consider only the events with the displaced vertex -- the $Z$ boson decays
\begin{equation}
    e^{+}+e^{-}\to Z, \quad Z \to N+\nu, \quad N \to l^{+}+l^{-}+\nu
\end{equation}

To summarize, we conclude that there is a complementarity between the mentioned signatures at colliders and the non-collider experiments. While the latter would probe mainly $d_{\alpha}$, the former may explore the couplings $d_{Z},d_{W}$. The displaced vertex signatures may contribute to the sensitivity only if $d_{Z,W}\neq 0$, since these couplings determine the production of the HNLs. In contrast, the missing energy signature may still provide the sensitivity, given that the HNL production, in this case, is also controlled by $d_{\alpha}$.

In addition, we should stress another complementarity -- between the lepton colliders may mainly probe the $d_{Z}$ coupling, the hadronic colliders suit better for probing $d_{W}$.

Similar to the hadron colliders case, HNLs may be additionally produced by decays of light mesons. However, such events would typically have a large multiplicity and do not have peculiar kinematic features which simplifies their distinguishing from the SM events. Therefore, further, we concentrate only on the production from $Z$ bosons.

\section{Hadron colliders}
\label{sec:hadronic-colliders}
\subsection{Background}
In~\cite{CMS:2022fut}, the search for HNLs with the mixing coupling has been performed using the statistics accumulated during 2016-2018, corresponding to the integrated luminosity $138\text{ fb}^{-1}$ at CMS. The results of this search may be extrapolated to the high-luminosity LHC, with the corresponding scaling of the SM background. 

To reduce backgrounds, the following selection cuts have been imposed:
\begin{itemize}
    \item[--] One prompt lepton $l_{1}$ and two displaced leptons $l_{2,3}$ within the pseudorapidity range $|\eta| <2.5$. 
    \item[--] Prompt electron (muon): $p_{T} > 30-32$ (25) GeV, transverse impact parameter $|d_{0}|<0.05 $ cm and longitudinal impact parameter $
    |d_{z}|<0.1\text{ cm}$.
    \item[--] Displaced electrons (muons): $p_{T} > 7$ (5) GeV, $|d_{0}| > 0.01$ cm, $|d_{z}| < 10\text{ cm}$. The total transverse momentum of the two displaced leptons should be $p_{T,23} > 15\text{ GeV}$.
    \item[--] The invariant mass of 3 leptons should be within $50\text{ GeV}<\sqrt{s_{123}}<80\text{ GeV}$; the invariant mass of the displaced leptons $\sqrt{s_{23}}$ should not be close to the invariant mass of the SM resonances (such as $\omega$, $\phi$, $J/\psi$,\dots).
    \item[--] Angular constraints: the angle between the HNL direction (assumed to be given by the vector from the primary vertex to the secondary vertex) and the direction given by the total momentum of $l_{2},p_{3}$ is  $\cos(\theta_{\text{SV},23}) > 0.99$; the azimuthal separation between the prompt and each of the displaced leptons should be $|\Delta \phi(l_{1},l_{2/3})| > 1$; the angular separation between $l_{2,3}$ should be $\Delta R(l_{2},l_{3}) = \sqrt{\Delta \eta_{23}^{2}+\Delta \phi_{23}^{2}} < 1$.
    \item[--] Maximal displacement constraints: displaced vertex within the tracker, i.e., the transverse distance $\Delta_{2D} < 0.5\text{ m}$ and the longitudinal distance $\Delta_{||}<3\text{ m}$.
\end{itemize}
The reconstruction efficiency for the prompt leptons is $\simeq 90\%$. The reconstruction efficiency for displaced leptons depends on the lepton type, its relative isolation, and the displacement. In particular, for the displacement 10 (25) cm, depending on the relative isolation, the efficiency for the electron reconstruction varies in the limits from 20\%-40\% to 60\%-80\% (15\%-20\% to 50\%-60\%). In contrast, for the muons with the displacement $10$ (50) cm, the numbers change from 85\%-90\% to 95\% (40\%-50\% to 80\%).

\begin{figure*}[!t]
    \centering
    \includegraphics[width=0.45\textwidth]{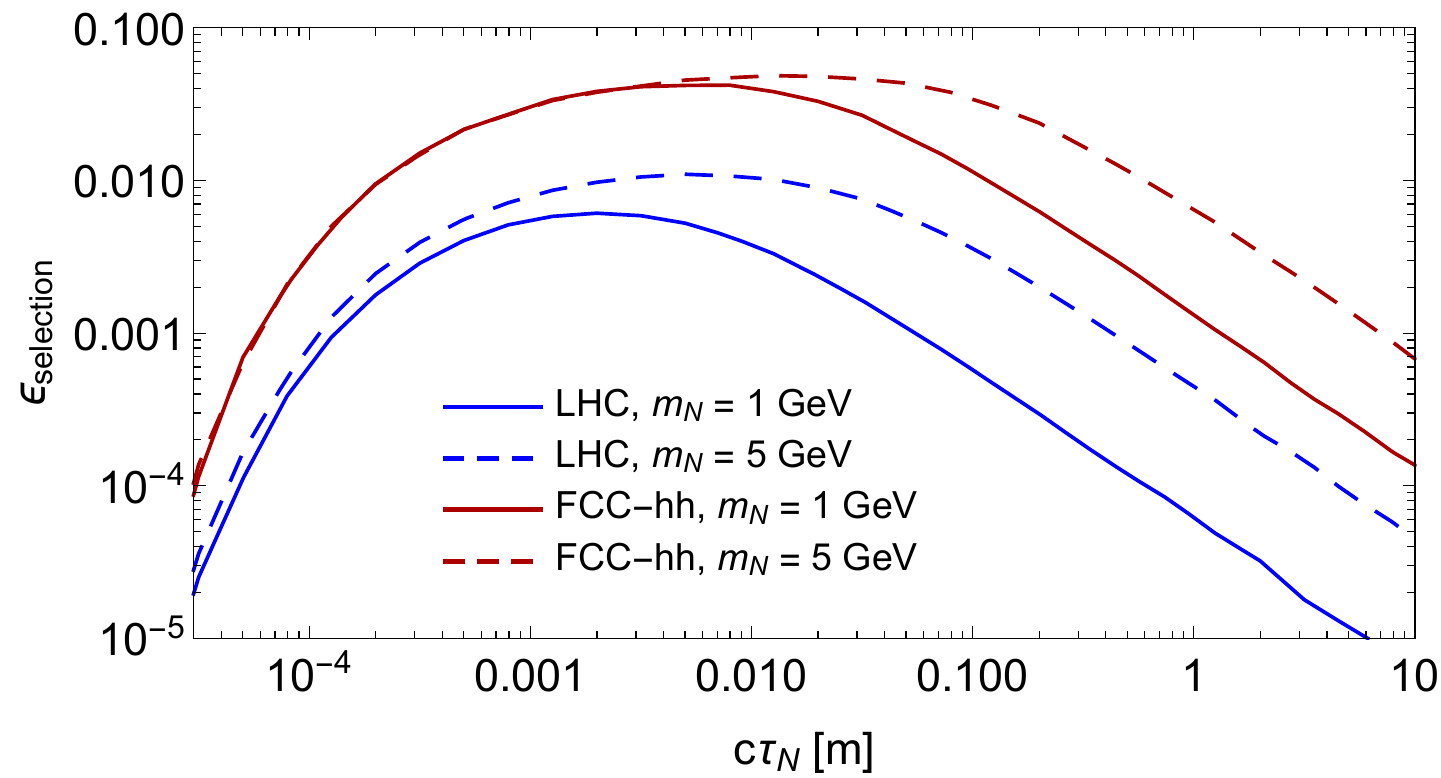}~\includegraphics[
    width=0.45\textwidth]{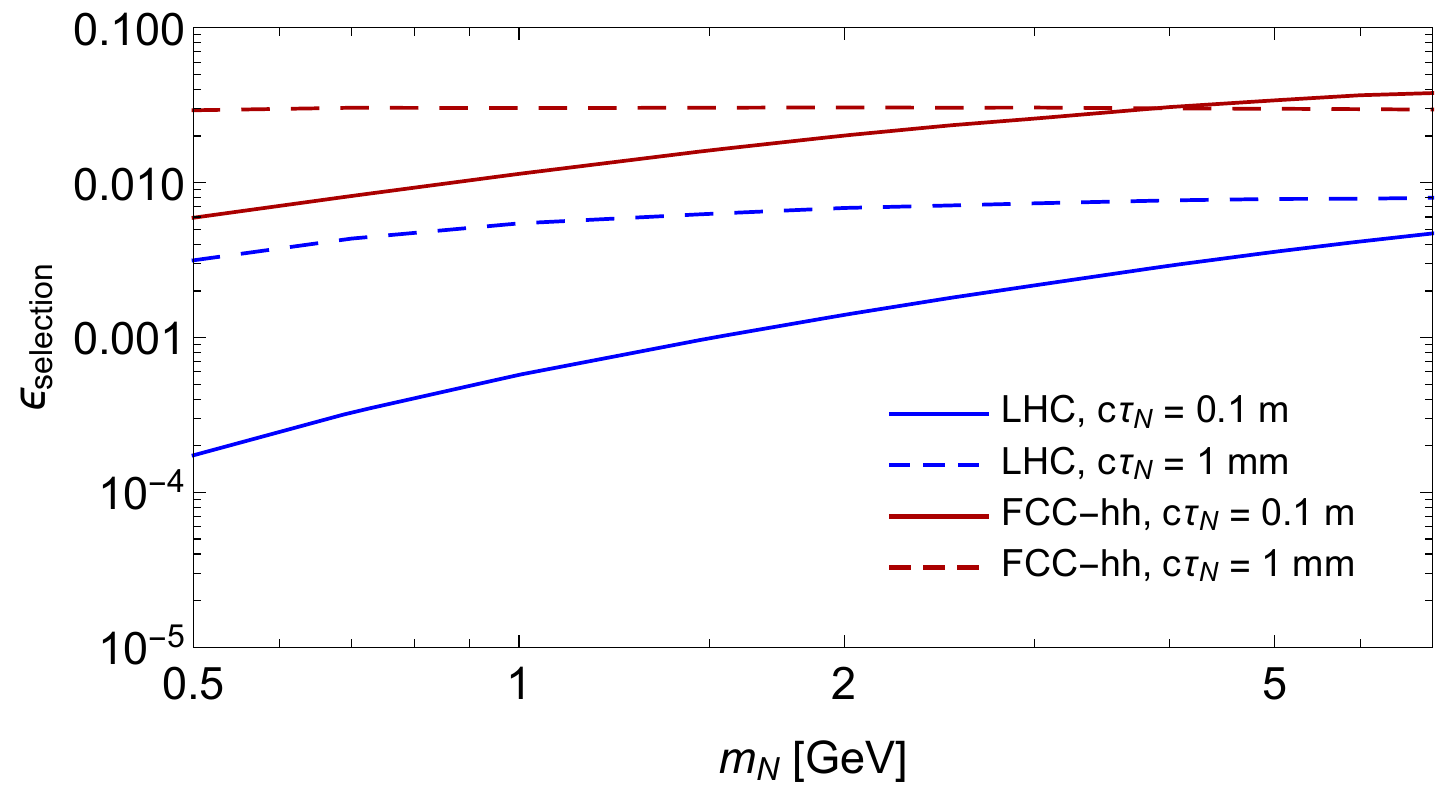}
    \caption{Selection efficiencies for the events with decaying HNLs at CMS@LHC and at FCC-hh, assuming the same experimental setup as at CMS@LHC. Left panel: as a function of the HNL decay length for several choices of its mass. Right panel: as a function of the HNL mass.}
    \label{fig:efficiency-hadronic}
\end{figure*}

Backgrounds for this selection set may come from the events with misidentified hadrons, muons from pion or kaon decays, and leptons coming from decays of heavy flavor hadrons. For the luminosity corresponding to the data set collected at CMS in 2016-2018, the total number of predicted background events is $\simeq 100-200$. The collected data agreed with the theoretical background prediction, which was used to impose the exclusion bound on the parameter space of the HNLs with the mixing coupling.

\subsection{Sensitivity}
\label{sec:hadronic-colliders-sensitivity}
Let us estimate the sensitivity of this scheme to the HNLs with the dipole coupling. We will consider the LHC in its high luminosity phase (HL-LHC) and FCC-hh, assuming for the latter the same search scheme as for the LHC. The parameters of these two detectors are summarized in Table~\ref{tab:hadronic-detector-parameters}. 
\begin{table}[ht!]
    \centering
    \begin{tabular}{|c|c|c|c|c|c|}
     \hline Detector & $|\eta|$ & $R\times L$ \\ \hline
      CMS@LHC  & $<2.5$  & $0.5\text{ m}\times 3\text{ m}$ \\ \hline FCC-hh & $<4$ & $1.6\text{ m}\times 5\text{ m}$ \\ \hline
    \end{tabular}
    \caption{Parameters of the trackers at CMS@LHC and the FCC-hh reference design detector: pseudorapidity coverage, transverse and longitudinal size. The values are taken from~\cite{Chatrchyan:2008aa} and~\cite{FCC:2018vvp}.}
    \label{tab:hadronic-detector-parameters}
\end{table}
Due to larger energies, the background at the FCC-hh may qualitatively change. Therefore, we will present the sensitivity of the FCC-hh in the form of iso-contours.

We start with evaluating the selection efficiency for the signal. We define it as
\begin{equation}
\epsilon_{\text{selection}} \equiv \frac{\sum_{l = e,\mu}\text{Br}(N\to \nu l^{+}l^{-})\times \epsilon_{\text{sel}}^{ll}}{\sum_{l = e,\mu}\text{Br}(N\to \nu l^{+}l^{-})},    
\end{equation}
where $\epsilon_{\text{sel}}^{ll}$ is the selection efficiency for the decay into a lepton pair $l^{+}l^{-}$. For simplicity, we perform a pure MC simulation, where the kinematics reconstruction effects are not considered. For the LHC, we approximate the displaced leptons reconstruction efficiency by a linear function of the transverse displacement, adopting conservatively the lowest values reported in~\cite{CMS:2022fut} for the interpolation points. As a cross-check of the calculations, we have reproduced the sensitivity to HNLs with the mixing coupling reported in~\cite{CMS:2022fut} within a factor of 1.5, which is appropriate given the simplicity of the simulation. For the FCC-hh, we assume unit displaced leptons reconstruction efficiency, motivated by a possible development of technologies at the time of the construction of FCC-hh. Compared to the CMS@LHC case, we also change the pseudorapidity/displacement cuts due to the changed tracker size (see Table~\ref{tab:hadronic-detector-parameters}), leaving the other cuts unchanged.

The mass and lifetime dependence of $\epsilon_{\text{selection}}$ for the HNLs with the dipole coupling case is shown in Fig.~\ref{fig:efficiency-hadronic}. From the figure, we see that for HNLs with mass $m_{N}\lesssim 10\text{ GeV}$, $\epsilon_{\text{selection}}$ does not exceed $\simeq 10^{-2}$ at the LHC. The corresponding values at the FCC-hh are at least one order of magnitude larger. This is a combined effect of the larger tracker volume and the unit displaced leptons reconstruction efficiency. For the fixed decay length, the efficiency increases with the HNL mass. The reason is an increase of the $p_{T}$ of the produced leptons relative to the direction of the incoming HNL, and hence the transverse impact parameter.

\begin{figure}[!h]
    \centering
    \includegraphics[width=0.45\textwidth]{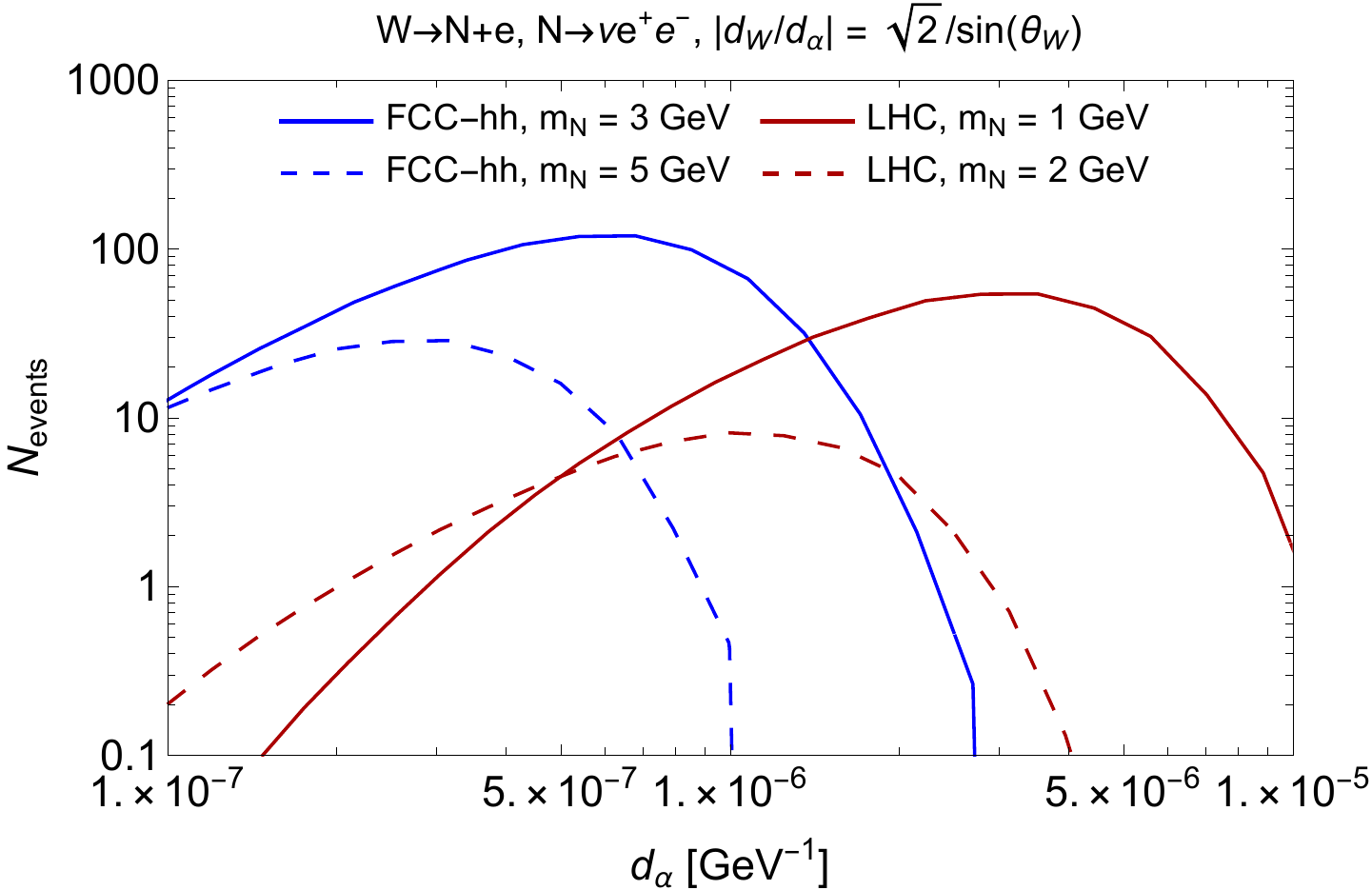}
    \includegraphics[width=0.45\textwidth]{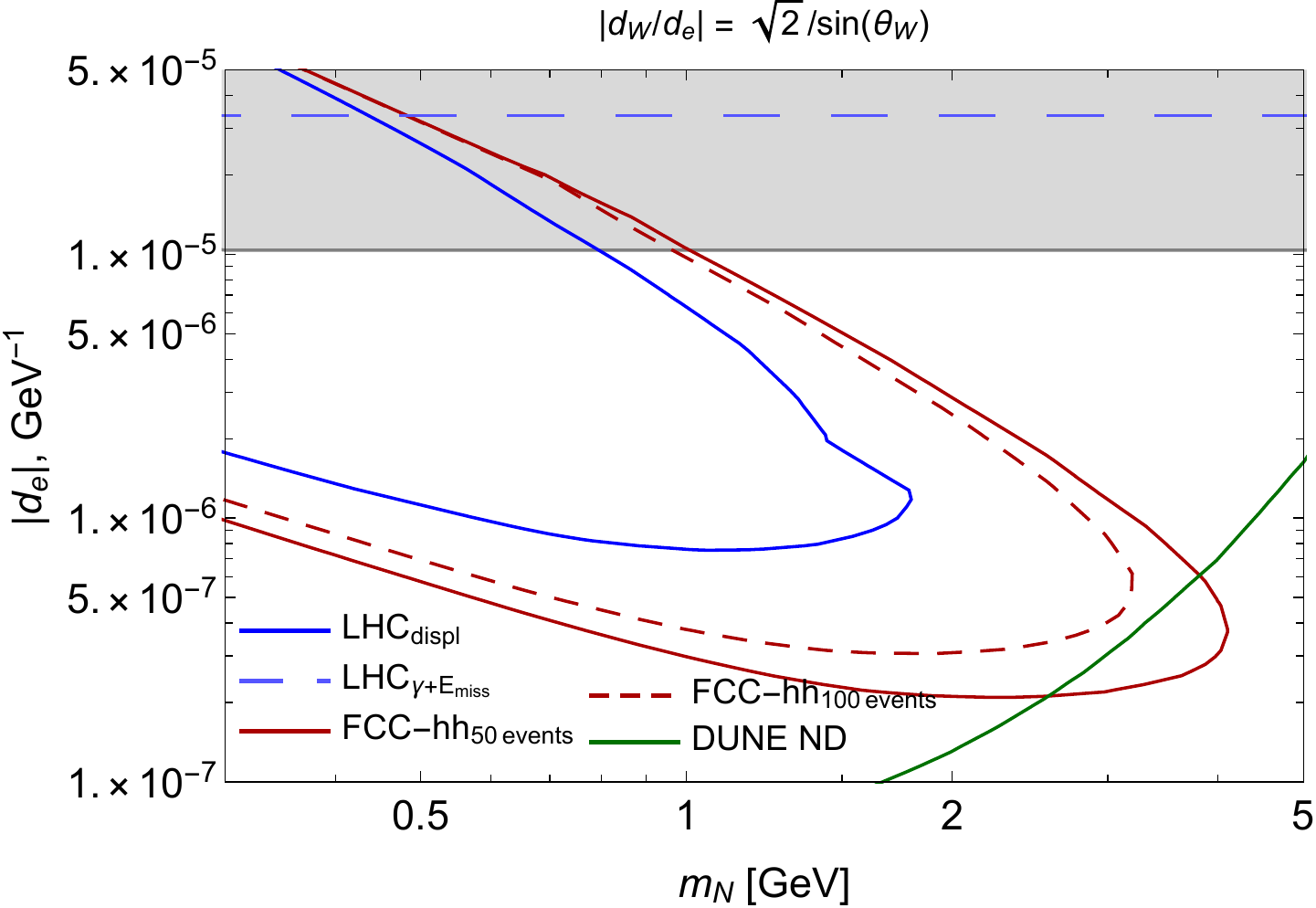}
    \caption{\textit{Left panel}: the behavior of the number of events for HNLs with different masses as a function of the dipole coupling $d_{\alpha}$. \textit{Right panel}: the potential of the hadron colliders -- high luminosity LHC and FCC-hh -- to probe the HNL parameter space. For the LHC case, we show the sensitivities coming from two signatures (Sec.~\ref{sec:HNLs-at-hadron-colliders}): the dilepton displaced vertex searches, for which we report the 90\% CL limit, as well as the projected sensitivity from the searches for the events with mono $\gamma$ and missing energy at ATLAS. In the case of the FCC-hh, we show the iso-contours corresponding to 
    50 and 100 events.}
    \label{fig:Nevents-hadronic-colliders}
\end{figure}

The number of events is given by
\begin{equation}
N_{\text{events}} = N_{W}\times \text{Br}(W\to N+l) \times\sum_{l = e,\mu}\text{Br}(N\to \nu l^{+}l^{-})\times \epsilon_{\text{selection}}    
\end{equation}
The behavior of the number of events with the coupling for the fixed mass is shown in Fig.~\ref{fig:Nevents-hadronic-colliders}. The number of events at the FCC-hh is a factor of a few hundred larger than at the LHC. This increase is due to the larger selection efficiency and a gain in the luminosity and $W$ boson production cross-section.

The sensitivities of the searches for the displaced vertices at the HL-LHC and FCC-hh assuming the coupling of the HNLs to the electron flavor are shown in Fig.~\ref{fig:Nevents-hadronic-colliders}. Although the sensitivity of the LHC is completely within the sensitivity of DUNE, it may still be a useful probe of the dipole portal, since it probes not only the $d_{\alpha}$ coupling, but also the coupling to $W$ bosons, and hence the LHC is complementary to other probes.

In the same figure, we also show the projected limits for the parameter space that may be probed by the mono $\gamma$ searches (remind Sec.~\ref{sec:HNLs-at-hadron-colliders}). Because of a huge background, this search cannot explore unconstrained HNL couplings.

\section{Lepton colliders}
\label{sec:lepton-colliders}
\subsection{Backgrounds}
\label{sec:backgrounds}
\begin{figure*}[!t]
    \centering
    \includegraphics[width=0.9\textwidth]{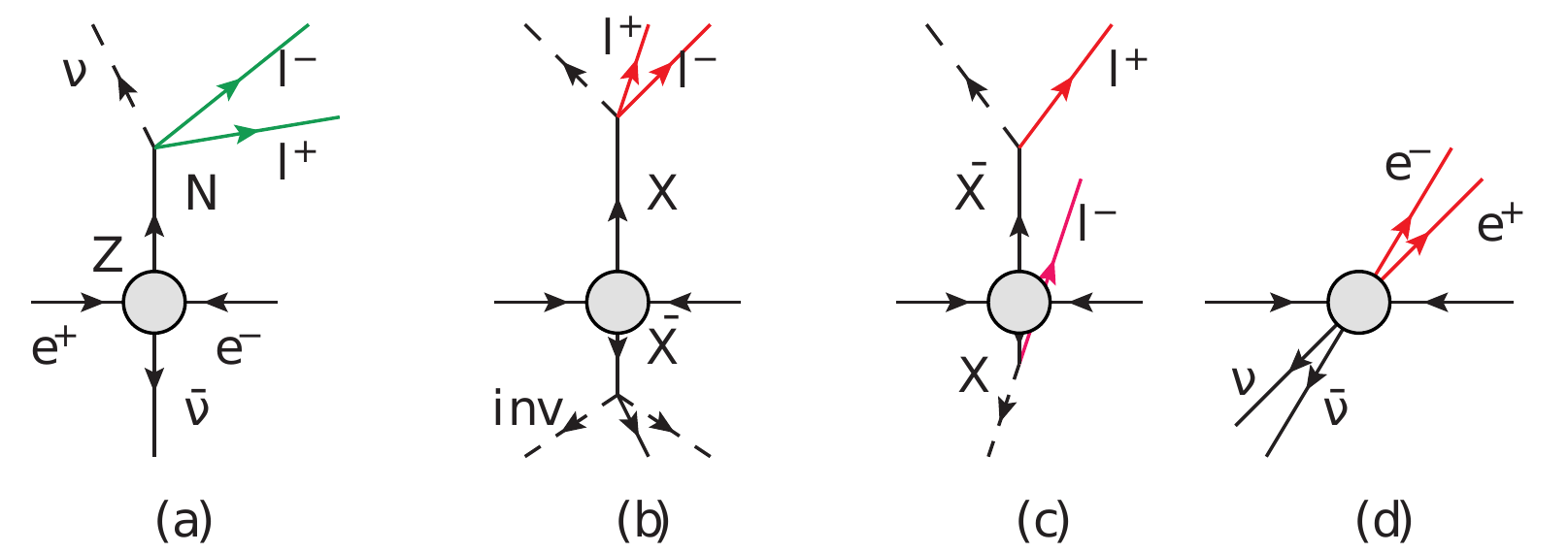}
    \caption{Events at lepton colliders. An event~\eqref{eq:signal} with an HNL decaying into a pair of charged leptons $l^{+},l^{-}$ (the diagram (a)), and possible background processes to it: decays $Z\to X\bar{X}\to l^{+}+l^{-}+\text{inv}$ (the diagrams (b), (c)), as well as 4-fermion process $e^{+}e^{-}\to l^{+}l^{-}+\nu+\bar{\nu}$.}
    \label{fig:events-topology}
\end{figure*}

Lepton colliders are free from pileup and have a low beam-induced background. Therefore, for the given process with an HNL, 
\begin{equation}
e^{+}+e^{-}\to Z \to N + \nu \to Y+\bar{Y}+\nu,
\label{eq:signal}
\end{equation}
where $Y,\bar{Y}$ denote visible HNL decay products,
the only possible background comes from single events of $e^{+},e^{-}$ collisions. The latter includes $Z$ boson decays
\begin{equation}
e^{+}+e^{-}\to Z\to f+\bar{f}\to Y+\bar{Y}+\text{inv},
\label{eq:bg-Z-decay}
\end{equation}
where $f = l = e,\mu,\tau$ or $q = u,d,s,c,b$, and prompt 4-fermion production
\begin{equation}
    e^{+}+e^{-}\to f +\bar{f}' + f'' + \bar{f}''' \to Y+\bar{Y}+\text{inv},
    \label{eq:bg-4-fermion}
\end{equation}
see Fig.~\ref{fig:events-topology}.

By ``inv'', we denote the particles that leave the detector invisibly; examples include neutrinos or the particles that have not been detected due to the inefficiency of the detector.

In~\cite{Blondel:2022qqo}, a preliminary background analysis for FCC-ee has been performed for the minimal HNL model with the mixing coupling. For the particular decay process $N\to e^{+}+e^{-}+\nu$, backgrounds from the decays of $Z$ bosons have been considered. The simulation started by generating events in MadGraph~\cite{Alwall:2014hca}, followed by Pythia8~\cite{Sjostrand:2014zea} for the hadronization and DELPHES~\cite{deFavereau:2013fsa} for the simulation of the detector response. The background reduction has been studied using pre-selection cuts, i.e. without requiring the candidates $Y$, $\bar{Y}$ to form a good vertex. The selection started from the requirement to have the visible final state consisting solely of a pair of $e^{+},e^{-}$ particles. Then, the event was required to have non-zero missing momentum $\slashed{p} = |\mathbf{p_{e^{+}}}+\mathbf{p_{e^{-}}}| > 10\text{ GeV}$, to account for finite momentum reconstruction resolution and remove a huge fraction of background from decays $Z\to ee$. Then, the cut on the transverse impact parameter, the minimal distance $|d_{0}|>0.5\text{ mm}$ from the track helical trajectory to the beamline, has been applied to both $e^{+}$ and $e^{-}$. This selection allowed reducing backgrounds from promptly produced $e^{+},e^{-}$. 

In total, the pre-selection reduced backgrounds down to $\sim 10^{5}$ - mostly coming from 
\begin{equation}
Z\to \tau+\bar{\tau} \to e^{+}+e^{-}+\text{inv}
\label{eq:background-tau-tau}
\end{equation}
Therefore, an additional selection is needed to remove the background. In addition, the impact parameter cut harms the sensitivity to short-lived HNLs, being in particular much more restrictive than the requirement for the vertex displacement $r_{\text{displ}}>400 \ \mu\text{m}$ used in~\cite{Blondel:2022qqo} to demonstrate the potential of FCC-ee to explore the parameter space of the HNLs with the mixing coupling (see also Fig.~\ref{fig:efficiency-lepton-colliders}).

An examination of the kinematics for the process~\eqref{eq:background-tau-tau} and the signal (remind Sec.~\ref{sec:hnl-phenomenology}) suggests that the amount of the remaining background events may be significantly reduced if imposing the cut on the angle between two electrons from above and their energies from below, see Appendix~\ref{app:additional-selection-cuts}. To study this question, we have performed a toy MC simulation of the process~\eqref{eq:background-tau-tau} and the events with HNLs in \texttt{Mathematica}. We have found that the cut 
\begin{equation}
    \cos(\theta_{ee}) > -0.5, \quad E_{e^{+}} > 2\text{ GeV}, \quad E_{e^{-}} > 2\text{ GeV}
    \label{eq:alternative-cuts}
\end{equation}
leaves no background events even before imposing the $|d_{0}|$ cut while keeping a large signal selection efficiency independent of the lifetime of the HNL. Apart from this selection, the background may also be reduced by requiring the electron-positron pair to form a good vertex (e.g., a small distance of the closest approach between their tracks). It may suggest that the $|d_{0}|$ cut can be relaxed to allow for probing short-lived HNLs, as the selection~\eqref{eq:alternative-cuts} should also work properly for the other $Z$ decays. A detailed simulation including the detector response is required to examine this question further, which is left for future work.

However, the cuts~\eqref{eq:alternative-cuts} are not efficient in the case of the 4-fermion production processes~\eqref{eq:bg-4-fermion}. To examine this question, we have simulated the purely leptonic process 
\begin{equation}
e^{+}+e^{-} \to e^{+}+e^{-}+\nu+\bar{\nu}
\end{equation}
in MadGraph. The total cross section of this process requiring $p_{T,l,\nu}>0.1\text{ GeV}$ has been found at the level of $\sigma_{ee\to ee\nu\nu}\approx 1.7\text{ pb}$, which results in $N_{Z}\cdot \frac{\sigma_{ee\to ee\nu\nu}}{\sigma_{ee\to Z}}\approx 2\cdot 10^{8}$ of such events during the $Z$-pole mode timeline. The $e^{+},e^{-}$ pair typically originates from the same vertex and hence may be as collimated as the signal, while neutrinos carry away missing momentum.

However, the 4-fermion process is prompt. Unlike the background coming from the decays of $Z$, the produced $e^{+},e^{-}$ pair has zero displacement from the collision point. To reduce this background to zero, one may additionally require non-zero displacement of the vertex formed by the $e^{+},e^{-}$ pair. The exact cut depends on the spatial resolution of the tracker. We will exploit two different choices for the displacement cut: 
\begin{equation}
r_{\text{displ}}>0.4 \ \text{mm}, \ \ \text{or} \ \  r_{\text{displ}} > 0.1\  \text{mm}
\end{equation}
The cuts considered in~\cite{Blondel:2022qqo} and the pre-selection we propose in this work are summarized in Table~\ref{tab:cuts-summary}.

\begin{table}[!h]
    \centering
    \begin{tabular}{|c|c|c|c|c|}
 \hline   & Selection cuts \\ \hline
    Ref.~\cite{Blondel:2022qqo} & \makecell{Only $e^{+},e^{-}$ in an event, $\slashed{p}>10\text{ GeV}$ \\ $|d_{0}|>0.5\text{ mm}$ } \\ \hline
  This work &  \makecell{Only $l^{+},l^{-}$ in an event, $\slashed{p} > 10\text{ GeV}$ \\ $\cos(\theta_{ll})>-0.5, \ E_{l^{+}},E_{l^{-}}>2\text{ GeV}$ \\ $r_{\text{displ}} >0.4\text{ mm}$,  or $r_{\text{displ}}>0.1\text{ mm}$} \\ \hline
    \end{tabular}
    \caption{Summary of the selection cuts required to remove the background for different HNL decay processes, as imposed in~\cite{Blondel:2022qqo} and considered in this work. Here, $d_{0}$ denotes the transverse impact parameter of any of the two tracks, $\slashed{p} = |\sum \mathbf{p}_{\text{reconstructed}}|$ corresponds to the missing momentum in an event, $\theta_{ab}$ is the angle between the two particles $a,b$, and $r_{\text{displ}}$ is the vertex displacement from the collision point.}
    \label{tab:cuts-summary}
\end{table}

\subsection{Sensitivity}
\label{sec:sensitivity}
\subsubsection{Selection efficiencies}
\begin{figure*}[!t]
    \centering
    \includegraphics[width=0.45\textwidth]{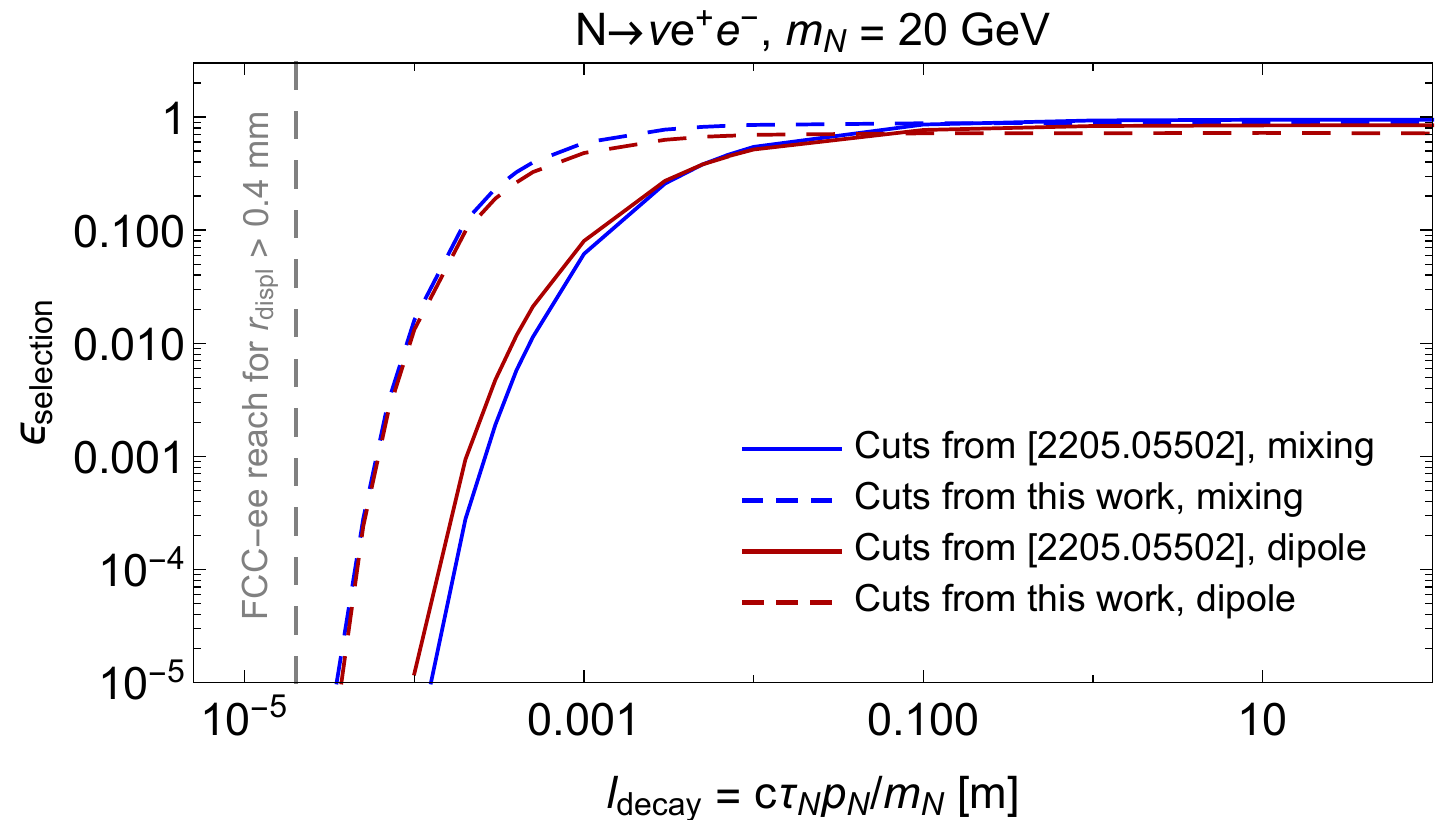}~\includegraphics[width=0.45\textwidth]{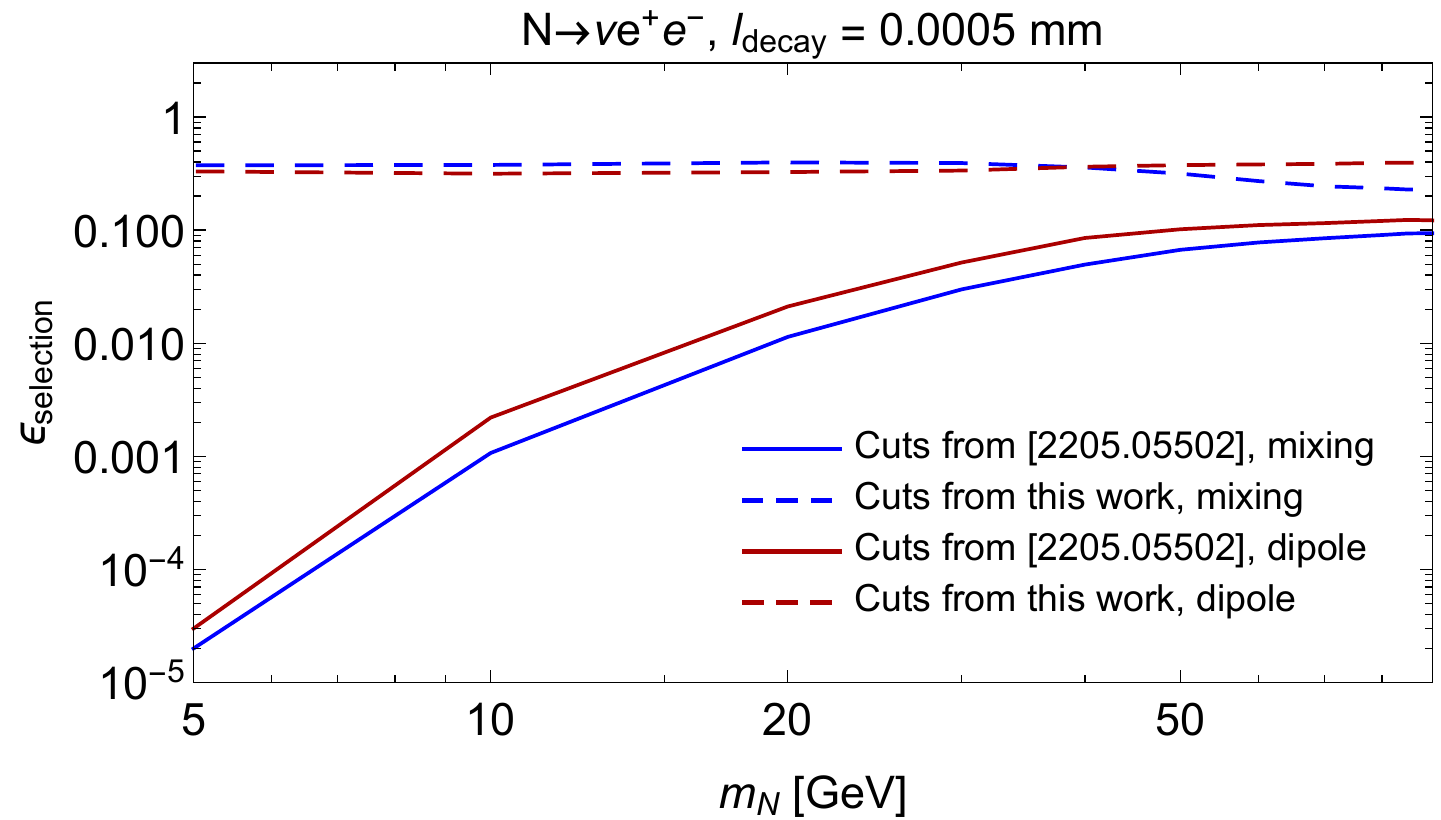}
    \caption{Selection efficiency for the process $N\to e^{+}e^{-}\nu$ (for both the mixing and dipole couplings) based on the cuts from Table~\ref{tab:cuts-summary}: the ones considered in~\cite{Blondel:2022qqo} (the blue lines), and the ones discussed in this work (the red lines), assuming the minimal displacement $l_{\text{displ}}>0.4\text{ mm}$. The left panel: as a function of the HNL decay length $l_{\text{N,decay}} = c\tau_{N}p_{N}/m_{N}$ for the fixed HNL mass $m_{N} = 30\text{ GeV}$. The vertical dashed gray line denotes the minimal decay length of the HNL with the mixing coupling to which FCC-ee may be sensitive if requiring only the displacement $r_{\text{displ}}>0.4\text{ mm}$ (from~\cite{Blondel:2022qqo}). The right panel: as a function of the HNL mass for the fixed HNL decay length $l_{\text{N,decay}} = 0.5$ mm.}
    \label{fig:efficiency-lepton-colliders}
\end{figure*}

The signal efficiency for the selection criteria from Table~\ref{tab:cuts-summary} for various HNL masses and decay lengths, considering both the mixing and dipole couplings, is shown in Fig.~\ref{fig:efficiency-lepton-colliders}. 

Let us first consider the cuts set from~\cite{Blondel:2022qqo}. We reproduce the values of the efficiencies reported for particular masses and lifetimes of HNLs with the mixing coupling in Table~3 of this paper. The figures show that the cuts' impact depends significantly on the HNL mass and lifetime. The efficiency, being $\approx 1$ for $l_{\text{decay}}\gg 0.5\text{ mm}$ independently on the HNL mass, starts dropping at $l_{\text{decay}} \simeq 1\text{ cm}$. For the given decay length, the decrease of $\epsilon$ is larger for smaller HNL masses. The reason is that the impact parameter (and hence the efficiency) of the decay products is higher if they gain large $p_{T}$ relatively to the direction of the HNL, and the magnitude of $p_{T}$ is controlled by the HNL mass. The efficiency for the dipole coupling case has similar behavior. However, the impact of efficiency however is less severe. Indeed, because of the kinematics of the decay process $N\to l^{+}l^{-}\nu$ (remind Sec.~\ref{sec:hnl-phenomenology}), in the dipole case, the leptons typically gain smaller energies and than in the mixing case. Due to this feature, their deflection relative to the HNL is larger, which results in a larger IP on average.

For the cuts set proposed in this paper, the situation is different. The decrease at small lifetimes is obviously less significant. As for the $m_{N}$ behavior, the efficiency slightly drops once mass increases because of an increase of the mean angle between leptons with $m_{N}$. In particular, for heavy HNLs with $m_{N}\simeq m_{Z}$ a sizable fraction of events may have $\cos(\theta) < -0.5$. This effect is more significant for HNLs with mixing because of the process's kinematics. On the other hand, since leptons produced via the dipole coupling are less energetic, the efficiency is lower at low HNL masses because of the $E_{l}$ cut.

\begin{figure}[!h]
    \centering
\includegraphics[width=0.7\textwidth]{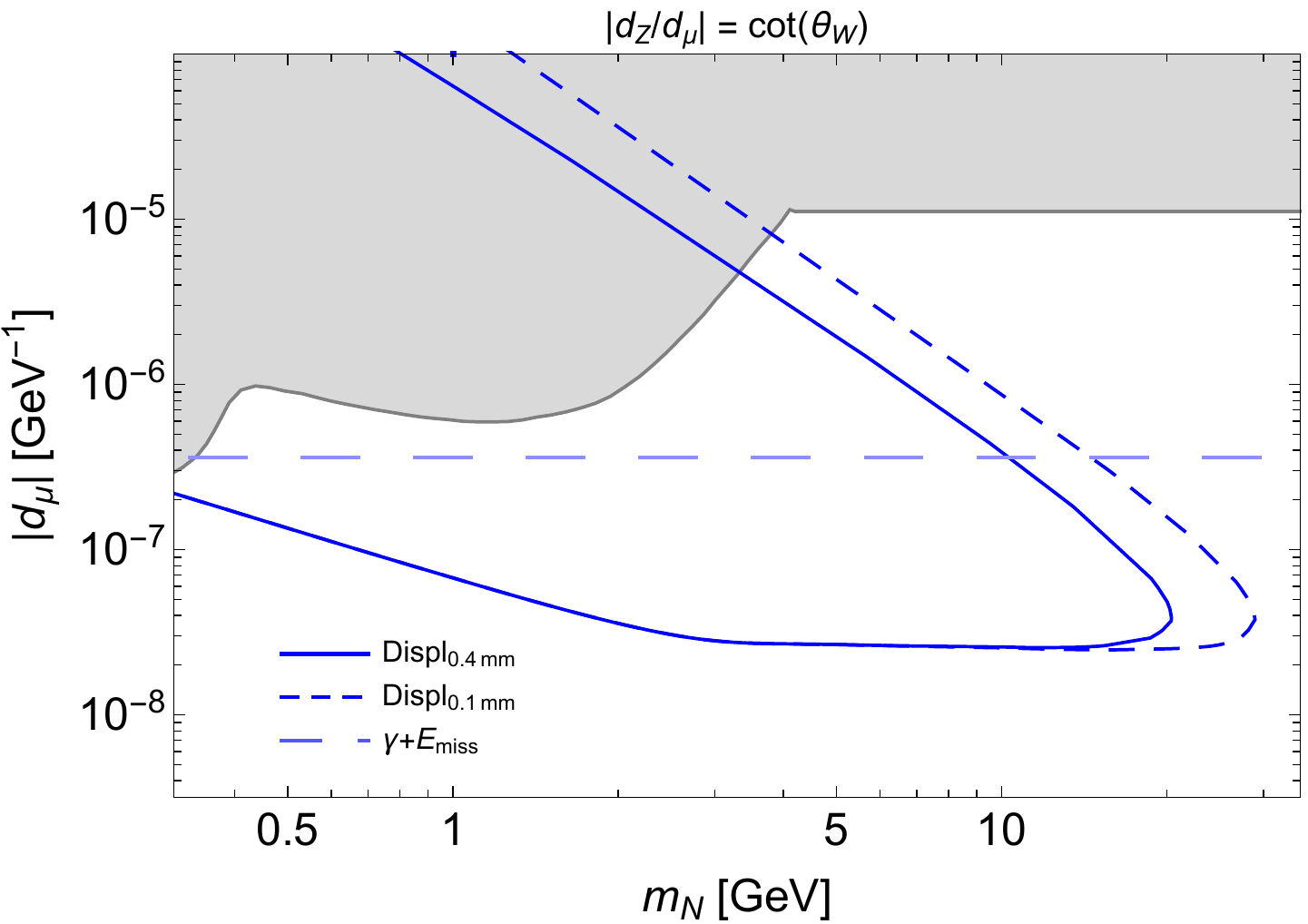}
    \caption{The potential of FCC-ee to probe the parameter space of the HNLs with the dipole coupling, see Sec.~\ref{sec:HNLs-at-lepton-colliders}. The solid and short-dashed dark blue lines show the 90\% CL sensitivity corresponding to the displaced decay signature, assuming the event selection considered in this paper (see Sec.~\ref{sec:backgrounds} and Table~\ref{tab:cuts-summary}). The long-dashed lighter blue line denotes the sensitivity corresponding to the $\gamma$+missing energy signature.}
    \label{fig:lepton-colliders-sensitivity}
\end{figure}

\subsubsection{Number of events and sensitivity curves}
Let us now estimate the sensitivity of FCC-ee to HNLs. We will consider the reference Innovative Detector for Electron–positron Accelerators (IDEA), which is a cylinder having the radius $r = 4.5\text{ m}$ and longitudinal size $L = 11\text{ m}$~\cite{Blondel:2022qqo}. The other reference detector, CLD, has very similar specifications, and therefore the sensitivity would be completely similar.

The expected number of events with decays of HNLs at IDEA@FCC-ee is
\begin{equation}
    N_{\text{events}} = 2\cdot N_{Z}\cdot\text{Br}_{Z\to N+\nu}\times \sum_{l = e,\mu}\text{Br}_{N\to l^{+}l^{-}\nu}\times \epsilon^{(l)}_{\text{sel}},
    \label{eq:nevents}
\end{equation}
where $\epsilon_{\text{sel}} = \epsilon_{\text{sel}}(m_{N},d_{\alpha})$ is the fraction of events with HNLs decaying inside the decay volume and that satisfy the selection cuts from Table~\ref{tab:cuts-summary}. In the limit when $l_{\text{decay,N}}\gg \mathcal{O}(1\text{ mm})$, the displacement selection has unit efficiency, and $\epsilon_{\text{sel}}$ becomes decay length-independent:
\begin{equation}
    \epsilon_{\text{sel}} \approx \frac{\epsilon(m_{N})}{\pi}\int \limits_{0}^{\pi}d\theta \left( \exp\left[-\frac{l_{\text{min}}}{l_{\text{decay,N}}} \right]-\exp\left[-\frac{l_{\text{max}}(\theta)}{l_{\text{decay,N}}} \right]\right),
\end{equation}
where the integration is performed over all directions of the cylindrical decay volume of IDEA.

The sensitivity of the FCC-ee to the HNLs with the dipole coupling is shown in Fig.~\ref{fig:lepton-colliders-sensitivity}, where we also include the sensitivity of the missing energy search (remind Sec.~\ref{sec:HNLs-at-lepton-colliders}). To fix the excluded parameter space, we assume $d_{\alpha} = d_{\mu}$. We stress however that the sensitivity of the FCC-ee is flavor-universal since both the production and decay of the HNL are flavor-agnostic. 

\begin{figure*}[!t]
    \centering
    \includegraphics[width=0.45\textwidth]{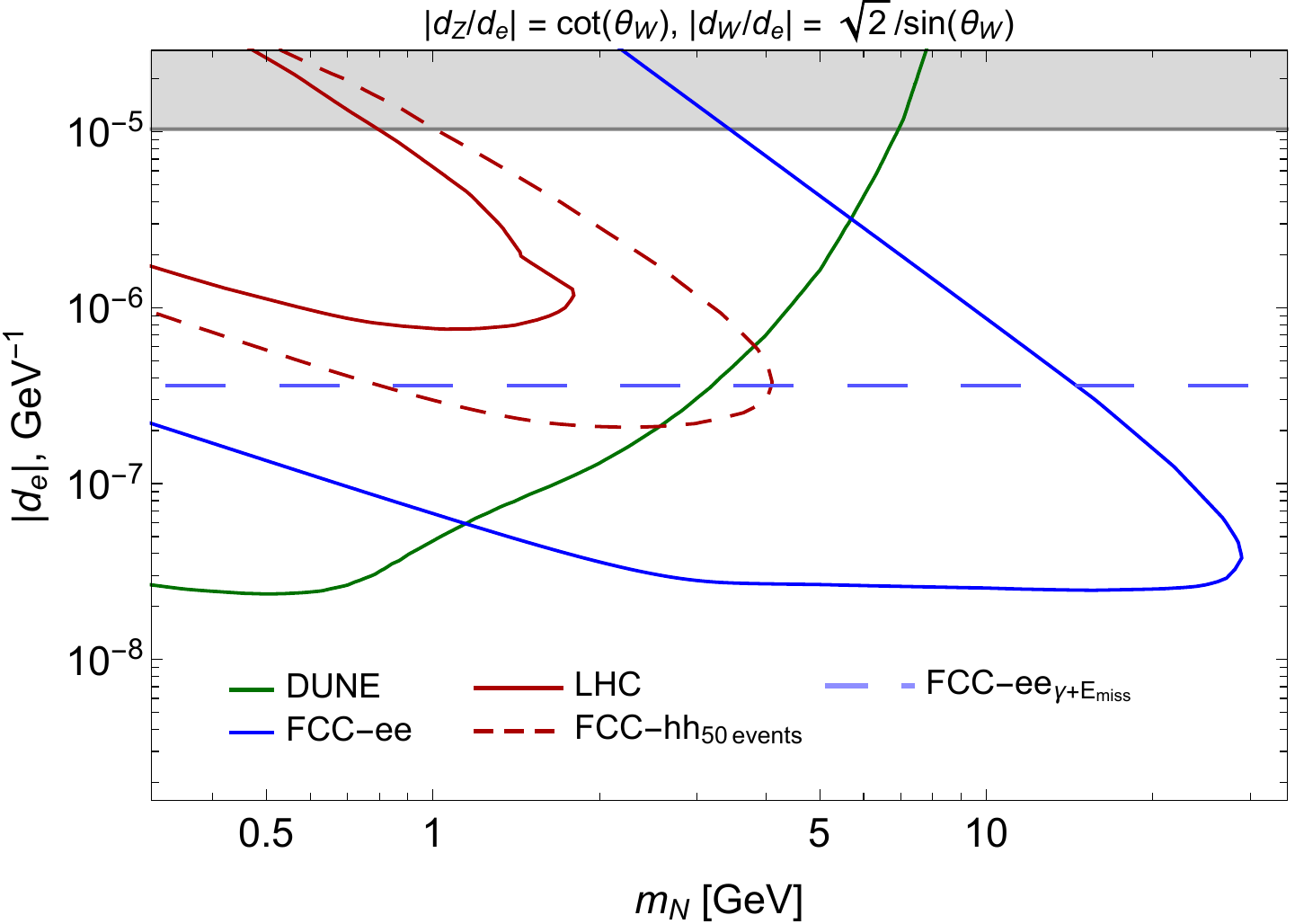}\\\includegraphics[width=0.45\textwidth]{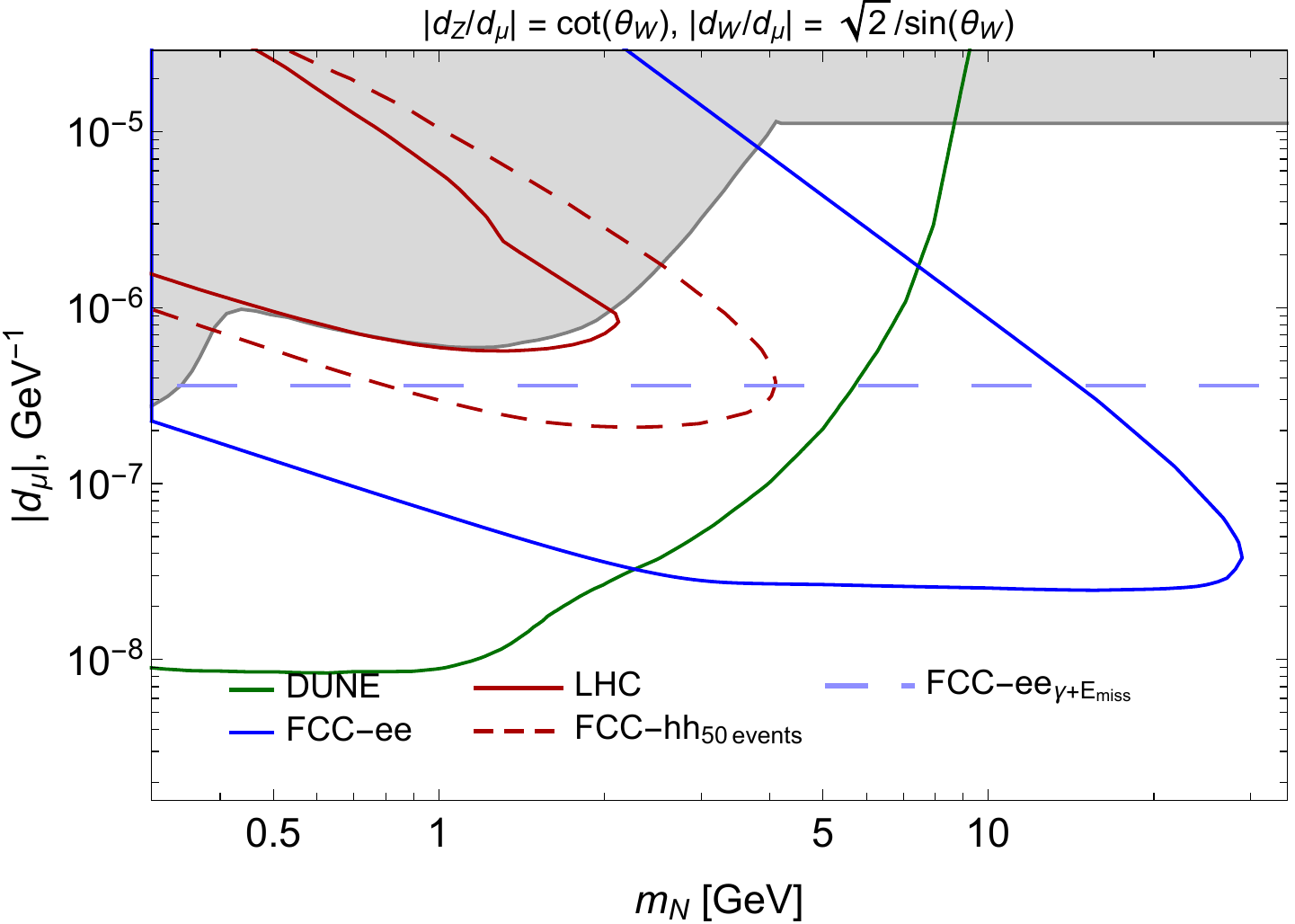}\includegraphics[width=0.45\textwidth]{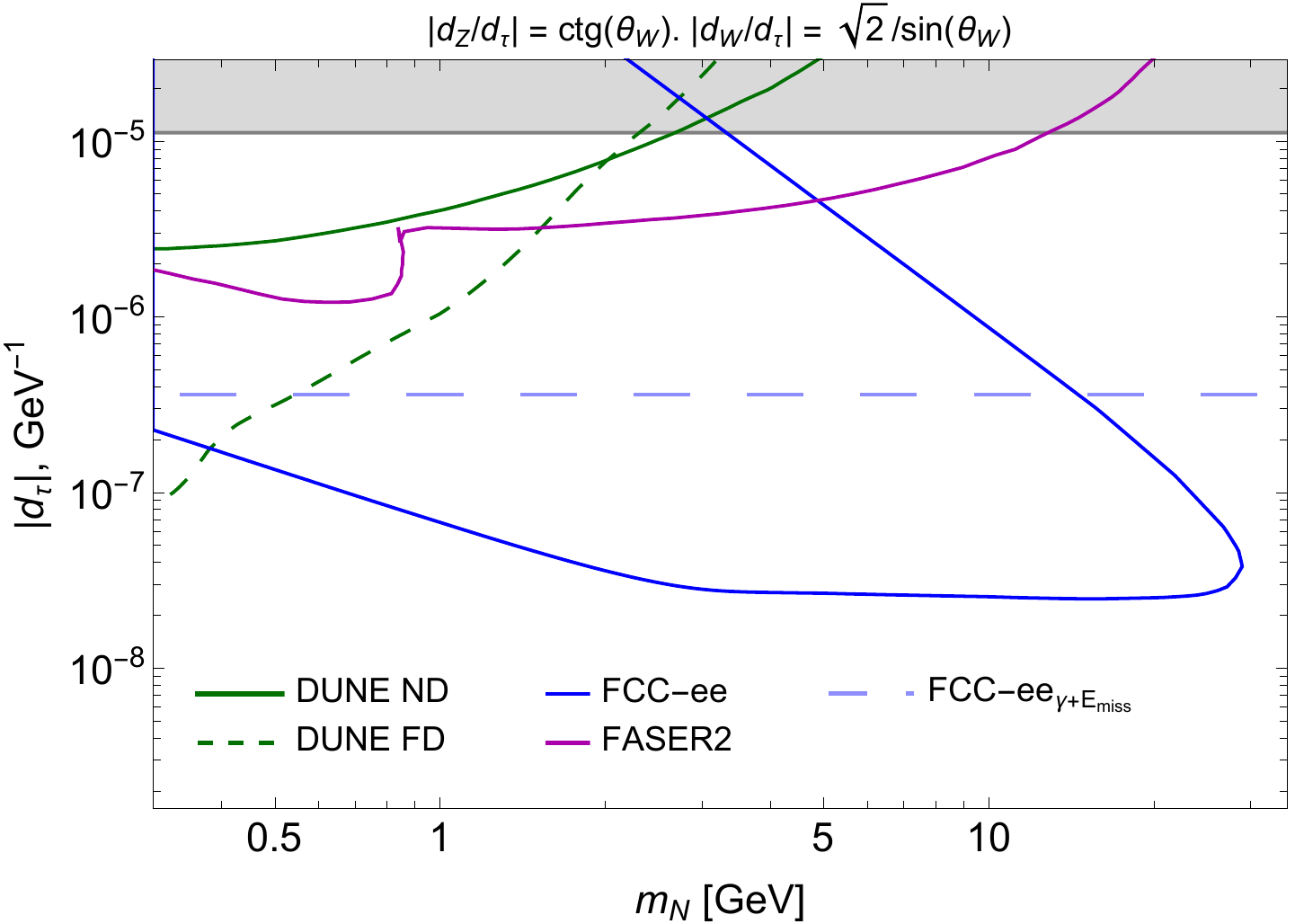}
    \caption{Potential of colliders -- FCC-ee, LHC in the high luminosity phase, and FCC-hh -- to explore the parameter space of HNLs with the dipole coupling. For the LHC, we report the 90\% CL sensitivity based on the search scheme and backgrounds from~\cite{CMS:2022fut} (see Sec.~\ref{sec:hadronic-colliders-sensitivity}). For FCC-hh, we assume the same search scheme as for the LHC and show the iso-contour corresponding to 50 events. For FCC-ee, we report the 90\% CL sensitivity assuming that the background is absent (see the corresponding discussion in Sec.~\ref{sec:backgrounds}).}
    \label{fig:final-sensitivity}
\end{figure*}

From the figure, we conclude that depending on the displacement cut, with the displaced decay searches FCC-ee may probe the HNLs with masses up to $m_{N} = 30$ GeV. The upper bound of the sensitivity is caused by the HNL decay vertex displacement selection. The shape of the lower bound is changing: below $m_{N}\simeq 3\text{ GeV}$ it gets smoothly improved, while at larger masses it becomes plateau. The reason is that at small masses, the HNL decay length at the lower bound is $l_{\text{N,decay}}\gg 1\text{ m}$, and therefore the decay probability scales as $P_{\text{decay}} \approx l_{\text{N,decay}}/l_{\text{fid}} \propto m_{N}^{-4}$, where the scaling comes from the behavior of the HNL decay width~\eqref{eq:decay-width} and the $\gamma$ factor. At large masses, the HNL decay length becomes small enough such that HNLs have a unit probability of decaying inside the detector. The lower bound in this case is determined by the condition $N_{\text{N,prod}}\times \epsilon_{\text{sel}} >2.3$, which is almost mass-independent in the mass range of interest.

The missing energy search is complementary compared to the displaced decays. Namely, it cannot probe as small couplings as probed by the displaced decay search because of significant background. However, it may explore higher HNL masses, since there is no displacement cut.

\section{Conclusions}
\label{sec:conclusions}

In this paper, we have analyzed the potential of hadron and lepton colliders to probe the parameter space of HNLs with the dipole coupling. 

We have first discussed the phenomenology of HNLs -- including their production, decays, and possible signatures -- at the LHC, FCC-hh, and FCC-ee (Sec.~\ref{sec:hnl-phenomenology}). We have also commented on how to distinguish decays of HNLs with mixing and dipole couplings. Thanks to the different working modes of the lepton and hadron colliders, they complement each other in exploring the parameter space of HNLs: the hadron colliders may probe the coupling of HNLs to $W$ bosons, while the lepton colliders are more efficient in probing the coupling to $Z$ bosons. In addition, because of the production channels of HNLs, from decays of $W,Z$ bosons, as well as due to the small distance from the production point to the decay volume, the colliders may probe the parameter space in the mass range inaccessible to neutrino factories such as DUNE and FASER2.

Then, we have considered the hadron colliders (Sec.~\ref{sec:hadronic-colliders}), 
utilizing the search for displaced vertices with dileptons at CMS as well as the missing energy searches at ATLAS. We have derived the sensitivity of the LHC in the high luminosity phase and estimated the potential of FCC-hh (Fig.~\ref{fig:Nevents-hadronic-colliders}). A detailed background study for the FCC-hh case is required, which however goes beyond the scope of this paper.

Next, we have considered the lepton colliders, see Sec.~\ref{sec:lepton-colliders}, concentrating on FCC-ee. We have first made a simplified background analysis demonstrating that the HNL decay signal can be promisingly distinguished from the background (Sec.~\ref{sec:backgrounds}), where we proposed new selection rules using kinematic properties. These findings will need to be checked with full-scale simulation in the future. Depending on the model parameters and given the ideal zero background, it may be possible to probe the HNL masses up to $m_{N}\simeq 30\text{ GeV}$, see Fig.~\ref{fig:lepton-colliders-sensitivity}. 

The final plot combining the sensitivities of lepton and hadron colliders is shown in Fig.~\ref{fig:final-sensitivity}, where we marginalize over the couplings to $Z,W$ assuming their maximal possible values. From the figures, we conclude that FCC-ee may explore the HNL masses up to $m_{N}\simeq 30 $ GeV, while the exploration potential of the hadron colliders is limited by $m_{N}\simeq 3$ GeV. This is due to the different background environments of these colliders: FCC-ee is free from pileup events, and therefore background is much cleaner, which allows for softer selection which keeps high efficiency for events with HNLs and simultaneously efficiently reduces the yield of the pure SM events. 

\section*{Acknowledgements}

We thank Juliette Alimena, Suchita Kulkarni, and Rebeca Gonzalez Suarez for discussing the background estimates at FCC-ee performed in~\cite{Blondel:2022qqo}, and Lesya Shchutska for discussing the backgrounds at the LHC. This project has received support from the European Union’s Horizon 2020 research and innovation program under the Marie Sklodowska-Curie grant agreement No. 860881-HIDDeN. Jing-yu Zhu is grateful for the support from the China and Germany Postdoctoral Exchange Program from the Office of China Postdoctoral Council and the Helmholtz Centre under Grant No. 2020031 and by the National Natural Science Foundation of China under Grant No. 11835005 and 11947227.


\bibliographystyle{JHEP}
\bibliography{bib}

\providecommand{\href}[2]{#2}\begingroup\raggedright\begin{thebibliography}{10}

\bibitem{Magill:2018jla}
G.~Magill, R.~Plestid, M.~Pospelov and Y.-D.~Tsai, \emph{{Dipole Portal to
  Heavy Neutral Leptons}},
  \href{https://doi.org/10.1103/PhysRevD.98.115015}{\emph{Phys. Rev. D}
  {\bfseries 98} (2018) 115015}
  [\href{https://arxiv.org/abs/1803.03262}{{\ttfamily 1803.03262}}].

\bibitem{Barducci:2022gdv}
D.~Barducci, E.~Bertuzzo, M.~Taoso and C.~Toni, \emph{{Probing right-handed
  neutrinos dipole operators}},
  \href{https://arxiv.org/abs/2209.13469}{{\ttfamily 2209.13469}}.

\bibitem{Brdar:2020quo}
V.~Brdar, A.~Greljo, J.~Kopp and T.~Opferkuch, \emph{{The Neutrino Magnetic
  Moment Portal: Cosmology, Astrophysics, and Direct Detection}},
  \href{https://doi.org/10.1088/1475-7516/2021/01/039}{\emph{JCAP} {\bfseries
  01} (2021) 039} [\href{https://arxiv.org/abs/2007.15563}{{\ttfamily
  2007.15563}}].

\bibitem{Gninenko:2009ks}
S.N.~Gninenko, \emph{{The MiniBooNE anomaly and heavy neutrino decay}},
  \href{https://doi.org/10.1103/PhysRevLett.103.241802}{\emph{Phys. Rev. Lett.}
  {\bfseries 103} (2009) 241802}
  [\href{https://arxiv.org/abs/0902.3802}{{\ttfamily 0902.3802}}].

\bibitem{Gninenko:2010pr}
S.N.~Gninenko, \emph{{A resolution of puzzles from the LSND, KARMEN, and
  MiniBooNE experiments}},
  \href{https://doi.org/10.1103/PhysRevD.83.015015}{\emph{Phys. Rev. D}
  {\bfseries 83} (2011) 015015}
  [\href{https://arxiv.org/abs/1009.5536}{{\ttfamily 1009.5536}}].

\bibitem{Shoemaker:2020kji}
I.M.~Shoemaker, Y.-D.~Tsai and J.~Wyenberg, \emph{{Active-to-sterile neutrino
  dipole portal and the XENON1T excess}},
  \href{https://doi.org/10.1103/PhysRevD.104.115026}{\emph{Phys. Rev. D}
  {\bfseries 104} (2021) 115026}
  [\href{https://arxiv.org/abs/2007.05513}{{\ttfamily 2007.05513}}].

\bibitem{Plestid:2020vqf}
R.~Plestid, \emph{{Luminous solar neutrinos I: Dipole portals}},
  \href{https://doi.org/10.1103/PhysRevD.104.075027}{\emph{Phys. Rev. D}
  {\bfseries 104} (2021) 075027}
  [\href{https://arxiv.org/abs/2010.04193}{{\ttfamily 2010.04193}}].

\bibitem{Jodlowski:2020vhr}
K.~Jod\l{}owski and S.~Trojanowski, \emph{{Neutrino beam-dump experiment with
  FASER at the LHC}},
  \href{https://doi.org/10.1007/JHEP05(2021)191}{\emph{JHEP} {\bfseries 05}
  (2021) 191} [\href{https://arxiv.org/abs/2011.04751}{{\ttfamily
  2011.04751}}].

\bibitem{Atkinson:2021rnp}
M.~Atkinson, P.~Coloma, I.~Martinez-Soler, N.~Rocco and I.M.~Shoemaker,
  \emph{{Heavy neutrino searches through double-bang events at
  Super-Kamiokande, DUNE, and Hyper-Kamiokande}},
  \href{https://doi.org/10.1007/JHEP04(2022)174}{\emph{JHEP} {\bfseries 04}
  (2022) 174} [\href{https://arxiv.org/abs/2105.09357}{{\ttfamily
  2105.09357}}].

\bibitem{Schwetz:2020xra}
T.~Schwetz, A.~Zhou and J.-Y.~Zhu, \emph{{Constraining active-sterile neutrino
  transition magnetic moments at DUNE near and far detectors}},
  \href{https://doi.org/10.1007/JHEP07(2021)200}{\emph{JHEP} {\bfseries 21}
  (2020) 200} [\href{https://arxiv.org/abs/2105.09699}{{\ttfamily
  2105.09699}}].

\bibitem{Dasgupta:2021fpn}
A.~Dasgupta, S.K.~Kang and J.E.~Kim, \emph{{Probing neutrino dipole portal at
  COHERENT experiment}},
  \href{https://doi.org/10.1007/JHEP11(2021)120}{\emph{JHEP} {\bfseries 11}
  (2021) 120} [\href{https://arxiv.org/abs/2108.12998}{{\ttfamily
  2108.12998}}].

\bibitem{Ismail:2021dyp}
A.~Ismail, S.~Jana and R.M.~Abraham, \emph{{Neutrino up-scattering via the
  dipole portal at forward LHC detectors}},
  \href{https://doi.org/10.1103/PhysRevD.105.055008}{\emph{Phys. Rev. D}
  {\bfseries 105} (2022) 055008}
  [\href{https://arxiv.org/abs/2109.05032}{{\ttfamily 2109.05032}}].

\bibitem{Miranda:2021kre}
O.G.~Miranda, D.K.~Papoulias, O.~Sanders, M.~T\'ortola and J.W.F.~Valle,
  \emph{{Low-energy probes of sterile neutrino transition magnetic moments}},
  \href{https://doi.org/10.1007/JHEP12(2021)191}{\emph{JHEP} {\bfseries 12}
  (2021) 191} [\href{https://arxiv.org/abs/2109.09545}{{\ttfamily
  2109.09545}}].

\bibitem{Bolton:2021pey}
P.D.~Bolton, F.F.~Deppisch, K.~Fridell, J.~Harz, C.~Hati and S.~Kulkarni,
  \emph{{Probing active-sterile neutrino transition magnetic moments with
  photon emission from CE\ensuremath{\nu}NS}},
  \href{https://doi.org/10.1103/PhysRevD.106.035036}{\emph{Phys. Rev. D}
  {\bfseries 106} (2022) 035036}
  [\href{https://arxiv.org/abs/2110.02233}{{\ttfamily 2110.02233}}].

\bibitem{Arguelles:2021dqn}
C.A.~Arg\"uelles, N.~Foppiani and M.~Hostert, \emph{{Heavy neutral leptons
  below the kaon mass at hodoscopic neutrino detectors}},
  \href{https://doi.org/10.1103/PhysRevD.105.095006}{\emph{Phys. Rev. D}
  {\bfseries 105} (2022) 095006}
  [\href{https://arxiv.org/abs/2109.03831}{{\ttfamily 2109.03831}}].

\bibitem{Mathur:2021trm}
V.~Mathur, I.M.~Shoemaker and Z.~Tabrizi, \emph{{Using DUNE to shed light on
  the electromagnetic properties of neutrinos}},
  \href{https://doi.org/10.1007/JHEP10(2022)041}{\emph{JHEP} {\bfseries 10}
  (2022) 041} [\href{https://arxiv.org/abs/2111.14884}{{\ttfamily
  2111.14884}}].

\bibitem{Li:2022bqr}
Y.-F.~Li and S.-y.~Xia, \emph{{Probing neutrino magnetic moments and the
  Xenon1T excess with coherent elastic solar neutrino scattering}},
  \href{https://doi.org/10.1103/PhysRevD.106.095022}{\emph{Phys. Rev. D}
  {\bfseries 106} (2022) 095022}
  [\href{https://arxiv.org/abs/2203.16525}{{\ttfamily 2203.16525}}].

\bibitem{Zhang:2022spf}
Y.~Zhang, M.~Song, R.~Ding and L.~Chen, \emph{{Neutrino dipole portal at
  electron colliders}},
  \href{https://doi.org/10.1016/j.physletb.2022.137116}{\emph{Phys. Lett. B}
  {\bfseries 829} (2022) 137116}
  [\href{https://arxiv.org/abs/2204.07802}{{\ttfamily 2204.07802}}].

\bibitem{Huang:2022pce}
G.-y.~Huang, S.~Jana, M.~Lindner and W.~Rodejohann, \emph{{Probing Heavy
  Sterile Neutrinos at Ultrahigh Energy Neutrino Telescopes via the Dipole
  Portal}},  \href{https://arxiv.org/abs/2204.10347}{{\ttfamily 2204.10347}}.

\bibitem{Gustafson:2022rsz}
R.A.~Gustafson, R.~Plestid and I.M.~Shoemaker, \emph{{Neutrino portals,
  terrestrial upscattering, and atmospheric neutrinos}},
  \href{https://doi.org/10.1103/PhysRevD.106.095037}{\emph{Phys. Rev. D}
  {\bfseries 106} (2022) 095037}
  [\href{https://arxiv.org/abs/2205.02234}{{\ttfamily 2205.02234}}].

\bibitem{Kamp:2022bpt}
N.W.~Kamp, M.~Hostert, A.~Schneider, S.~Vergani, C.A.~Arg\"uelles, J.M.~Conrad
  et~al., \emph{{Dipole-Coupled Neutrissimo Explanations of the MiniBooNE
  Excess Including Constraints from MINERvA Data}},
  \href{https://arxiv.org/abs/2206.07100}{{\ttfamily 2206.07100}}.

\bibitem{Abdullahi:2022cdw}
A.M.~Abdullahi, J.~Hoefken~Zink, M.~Hostert, D.~Massaro and S.~Pascoli,
  \emph{{DarkNews: a Python-based event generator for heavy neutral lepton
  production in neutrino-nucleus scattering}},
  \href{https://arxiv.org/abs/2207.04137}{{\ttfamily 2207.04137}}.

\bibitem{Delgado:2022fea}
F.~Delgado, L.~Duarte, J.~Jones-Perez, C.~Manrique-Chavil and S.~Pe\~na,
  \emph{{Assessment of the dimension-5 seesaw portal and impact of exotic Higgs
  decays on non-pointing photon searches}},
  \href{https://doi.org/10.1007/JHEP09(2022)079}{\emph{JHEP} {\bfseries 09}
  (2022) 079} [\href{https://arxiv.org/abs/2205.13550}{{\ttfamily
  2205.13550}}].

\bibitem{Ovchynnikov:2022rqj}
M.~Ovchynnikov, T.~Schwetz and J.-Y.~Zhu, \emph{{Dipole portal and
  neutrinophilic scalars at DUNE revisited: the importance of the high-energy
  neutrino tail}},  \href{https://arxiv.org/abs/2210.13141}{{\ttfamily
  2210.13141}}.

\bibitem{Abdullahi:2022jlv}
A.M.~Abdullahi et~al., \emph{{The Present and Future Status of Heavy Neutral
  Leptons}},  in \emph{{2022 Snowmass Summer Study}}, 3, 2022
  [\href{https://arxiv.org/abs/2203.08039}{{\ttfamily 2203.08039}}].

\bibitem{Zhang:2023nxy}
Y.~Zhang and W.~Liu, \emph{{Probing active-sterile neutrino transition magnetic
  moments at LEP and CEPC}},
  \href{https://arxiv.org/abs/2301.06050}{{\ttfamily 2301.06050}}.

\bibitem{SNDLHC:2022ihg}
{\scshape SND@LHC} collaboration, \emph{{SND@LHC: The Scattering and Neutrino
  Detector at the LHC}},  \href{https://arxiv.org/abs/2210.02784}{{\ttfamily
  2210.02784}}.

\bibitem{FASER:2020gpr}
{\scshape FASER} collaboration, \emph{{Technical Proposal: FASERnu}},
  \href{https://arxiv.org/abs/2001.03073}{{\ttfamily 2001.03073}}.

\bibitem{Feng:2022inv}
J.L.~Feng et~al., \emph{{The Forward Physics Facility at the High-Luminosity
  LHC}},  \href{https://arxiv.org/abs/2203.05090}{{\ttfamily 2203.05090}}.

\bibitem{SHiP:2015vad}
{\scshape SHiP} collaboration, \emph{{A facility to Search for Hidden Particles
  (SHiP) at the CERN SPS}},  \href{https://arxiv.org/abs/1504.04956}{{\ttfamily
  1504.04956}}.

\bibitem{DUNE:2020lwj}
{\scshape DUNE} collaboration, \emph{{Deep Underground Neutrino Experiment
  (DUNE), Far Detector Technical Design Report, Volume I Introduction to
  DUNE}}, \href{https://doi.org/10.1088/1748-0221/15/08/T08008}{\emph{JINST}
  {\bfseries 15} (2020) T08008}
  [\href{https://arxiv.org/abs/2002.02967}{{\ttfamily 2002.02967}}].

\bibitem{FCC:2018byv}
{\scshape FCC} collaboration, \emph{{FCC Physics Opportunities}: {Future
  Circular Collider Conceptual Design Report Volume 1}},
  \href{https://doi.org/10.1140/epjc/s10052-019-6904-3}{\emph{Eur. Phys. J. C}
  {\bfseries 79} (2019) 474}.

\bibitem{CEPCStudyGroup:2018rmc}
{\scshape CEPC Study Group} collaboration, \emph{{CEPC Conceptual Design
  Report: Volume 1 - Accelerator}},
  \href{https://arxiv.org/abs/1809.00285}{{\ttfamily 1809.00285}}.

\bibitem{CEPCStudyGroup:2018ghi}
{\scshape CEPC Study Group} collaboration, \emph{{CEPC Conceptual Design
  Report: Volume 2 - Physics \& Detector}},
  \href{https://arxiv.org/abs/1811.10545}{{\ttfamily 1811.10545}}.

\bibitem{Aime:2022flm}
C.~Aime et~al., \emph{{Muon Collider Physics Summary}},
  \href{https://arxiv.org/abs/2203.07256}{{\ttfamily 2203.07256}}.

\bibitem{Black:2022cth}
K.M.~Black et~al., \emph{{Muon Collider Forum Report}},
  \href{https://arxiv.org/abs/2209.01318}{{\ttfamily 2209.01318}}.

\bibitem{FCC:2018vvp}
{\scshape FCC} collaboration, \emph{{FCC-hh: The Hadron Collider}: {Future
  Circular Collider Conceptual Design Report Volume 3}},
  \href{https://doi.org/10.1140/epjst/e2019-900087-0}{\emph{Eur. Phys. J. ST}
  {\bfseries 228} (2019) 755}.

\bibitem{ATLAS:2016fij}
{\scshape ATLAS} collaboration, \emph{{Measurement of $W^{\pm}$ and $Z$-boson
  production cross sections in $pp$ collisions at $\sqrt{s}=13$ TeV with the
  ATLAS detector}},
  \href{https://doi.org/10.1016/j.physletb.2016.06.023}{\emph{Phys. Lett. B}
  {\bfseries 759} (2016) 601}
  [\href{https://arxiv.org/abs/1603.09222}{{\ttfamily 1603.09222}}].

\bibitem{Alwall:2014hca}
J.~Alwall, R.~Frederix, S.~Frixione, V.~Hirschi, F.~Maltoni, O.~Mattelaer
  et~al., \emph{{The automated computation of tree-level and next-to-leading
  order differential cross sections, and their matching to parton shower
  simulations}}, \href{https://doi.org/10.1007/JHEP07(2014)079}{\emph{JHEP}
  {\bfseries 07} (2014) 079} [\href{https://arxiv.org/abs/1405.0301}{{\ttfamily
  1405.0301}}].

\bibitem{ATLAS:2017nga}
{\scshape ATLAS} collaboration, \emph{{Search for dark matter at $\sqrt{s}=13$
  TeV in final states containing an energetic photon and large missing
  transverse momentum with the ATLAS detector}},
  \href{https://doi.org/10.1140/epjc/s10052-017-4965-8}{\emph{Eur. Phys. J. C}
  {\bfseries 77} (2017) 393}
  [\href{https://arxiv.org/abs/1704.03848}{{\ttfamily 1704.03848}}].

\bibitem{CMS:2022fut}
{\scshape CMS} collaboration, \emph{{Search for long-lived heavy neutral
  leptons with displaced vertices in proton-proton collisions at $
  \sqrt{\mathrm{s}} $ =13 TeV}},
  \href{https://doi.org/10.1007/JHEP07(2022)081}{\emph{JHEP} {\bfseries 07}
  (2022) 081} [\href{https://arxiv.org/abs/2201.05578}{{\ttfamily
  2201.05578}}].

\bibitem{Kling:2021fwx}
F.~Kling and S.~Trojanowski, \emph{{Forward experiment sensitivity estimator
  for the LHC and future hadron colliders}},
  \href{https://doi.org/10.1103/PhysRevD.104.035012}{\emph{Phys. Rev. D}
  {\bfseries 104} (2021) 035012}
  [\href{https://arxiv.org/abs/2105.07077}{{\ttfamily 2105.07077}}].

\bibitem{L3:1992cmn}
{\scshape L3} collaboration, \emph{{Search for anomalous production of single
  photon events in e+ e- annihilations at the Z resonance}},
  \href{https://doi.org/10.1016/0370-2693(92)91286-I}{\emph{Phys. Lett. B}
  {\bfseries 297} (1992) 469}.

\bibitem{OPAL:1994kgw}
{\scshape OPAL} collaboration, \emph{{Measurement of single photon production
  in e+ e- collisions near the Z0 resonance}},
  \href{https://doi.org/10.1007/BF01571303}{\emph{Z. Phys. C} {\bfseries 65}
  (1995) 47}.

\bibitem{DELPHI:1996drf}
{\scshape DELPHI} collaboration, \emph{{Search for new phenomena using single
  photon events in the DELPHI detector at LEP}},
  \href{https://doi.org/10.1007/s002880050421}{\emph{Z. Phys. C} {\bfseries 74}
  (1997) 577}.

\bibitem{Bondarenko:2018ptm}
K.~Bondarenko, A.~Boyarsky, D.~Gorbunov and O.~Ruchayskiy, \emph{{Phenomenology
  of GeV-scale Heavy Neutral Leptons}},
  \href{https://doi.org/10.1007/JHEP11(2018)032}{\emph{JHEP} {\bfseries 11}
  (2018) 032} [\href{https://arxiv.org/abs/1805.08567}{{\ttfamily
  1805.08567}}].

\bibitem{Ding:2019tqq}
J.-N.~Ding, Q.~Qin and F.-S.~Yu, \emph{{Heavy neutrino searches at future
  $Z$-factories}},
  \href{https://doi.org/10.1140/epjc/s10052-019-7277-3}{\emph{Eur. Phys. J. C}
  {\bfseries 79} (2019) 766}
  [\href{https://arxiv.org/abs/1903.02570}{{\ttfamily 1903.02570}}].

\bibitem{Shen:2022ffi}
Y.-F.~Shen, J.-N.~Ding and Q.~Qin, \emph{{Monojet search for heavy neutrinos at
  future Z-factories}},
  \href{https://doi.org/10.1140/epjc/s10052-022-10301-4}{\emph{Eur. Phys. J. C}
  {\bfseries 82} (2022) 398}
  [\href{https://arxiv.org/abs/2201.05831}{{\ttfamily 2201.05831}}].

\bibitem{An:2018dwb}
F.~An et~al., \emph{{Precision Higgs physics at the CEPC}},
  \href{https://doi.org/10.1088/1674-1137/43/4/043002}{\emph{Chin. Phys. C}
  {\bfseries 43} (2019) 043002}
  [\href{https://arxiv.org/abs/1810.09037}{{\ttfamily 1810.09037}}].

\bibitem{Boscolo:2019awb}
M.~Boscolo et~al., \emph{{Machine detector interface for the $e^+e^-$ future
  circular collider}},  in \emph{{62nd ICFA Advanced Beam Dynamics Workshop on
  High Luminosity Circular $e^+ e^-$ Colliders}}, p.~WEXBA02, 2019,
  \href{https://doi.org/10.18429/JACoW-eeFACT2018-WEXBA02}{DOI}
  [\href{https://arxiv.org/abs/1905.03528}{{\ttfamily 1905.03528}}].

\bibitem{Chatrchyan:2008aa}
{\scshape CMS} collaboration, \emph{{The CMS Experiment at the CERN LHC}},
  \href{https://doi.org/10.1088/1748-0221/3/08/S08004}{\emph{JINST} {\bfseries
  3} (2008) S08004}.

\bibitem{Blondel:2022qqo}
A.~Blondel et~al., \emph{{Searches for long-lived particles at the future
  FCC-ee}}, \href{https://doi.org/10.3389/fphy.2022.967881}{\emph{Front. in
  Phys.} {\bfseries 10} (2022) 967881}
  [\href{https://arxiv.org/abs/2203.05502}{{\ttfamily 2203.05502}}].

\bibitem{Sjostrand:2014zea}
T.~Sj\"ostrand, S.~Ask, J.R.~Christiansen, R.~Corke, N.~Desai, P.~Ilten et~al.,
  \emph{{An introduction to PYTHIA 8.2}},
  \href{https://doi.org/10.1016/j.cpc.2015.01.024}{\emph{Comput. Phys. Commun.}
  {\bfseries 191} (2015) 159}
  [\href{https://arxiv.org/abs/1410.3012}{{\ttfamily 1410.3012}}].

\bibitem{deFavereau:2013fsa}
{\scshape DELPHES 3} collaboration, \emph{{DELPHES 3, A modular framework for
  fast simulation of a generic collider experiment}},
  \href{https://doi.org/10.1007/JHEP02(2014)057}{\emph{JHEP} {\bfseries 02}
  (2014) 057} [\href{https://arxiv.org/abs/1307.6346}{{\ttfamily 1307.6346}}].

\bibitem{Workman:2022ynf}
{\scshape Particle Data Group} collaboration, \emph{{Review of Particle
  Physics}}, \href{https://doi.org/10.1093/ptep/ptac097}{\emph{PTEP} {\bfseries
  2022} (2022) 083C01}.

\end{thebibliography}\endgroup

\newpage

\appendix

\section{Events selection at FCC-ee}
\label{app:additional-selection-cuts}

Let us first analyze the kinematics of $Z$ boson decays into two $\tau$ leptons at FCC-ee. Since $Z$s are at rest, their decay products $\tau,\bar{\tau}$ fly in exactly opposite directions and have the same energy $E_{\tau} = E_{\bar{\tau}} = m_{Z}/2$. The $e^{+},e^{-}$ pair without any other visible particle can originate only from the two decays (the diagram (c) in Fig.~\ref{fig:events-topology})
\begin{equation}
\tau \to e^{-}+\bar{\nu}_{e}+\nu_{\tau},
\quad \bar{\tau} \to e^{+}+\nu_{e}+\bar{\nu}_{\tau}\;,
\end{equation}
where the distribution of $e^{+},e^{-}$ in the angle $\theta_{ee}$ between their directions of motion is peaked around $\theta_{ee} = \pi$. A small fraction of events with a small angle between the momenta of $e^{+},e^{-}$ have the following pattern: one of the particles from the pair has very small energy, $E_{e^{\pm}}\ll m_{Z}/2$. 

\begin{figure}[!h]
    \centering
\includegraphics[width=0.7\textwidth]{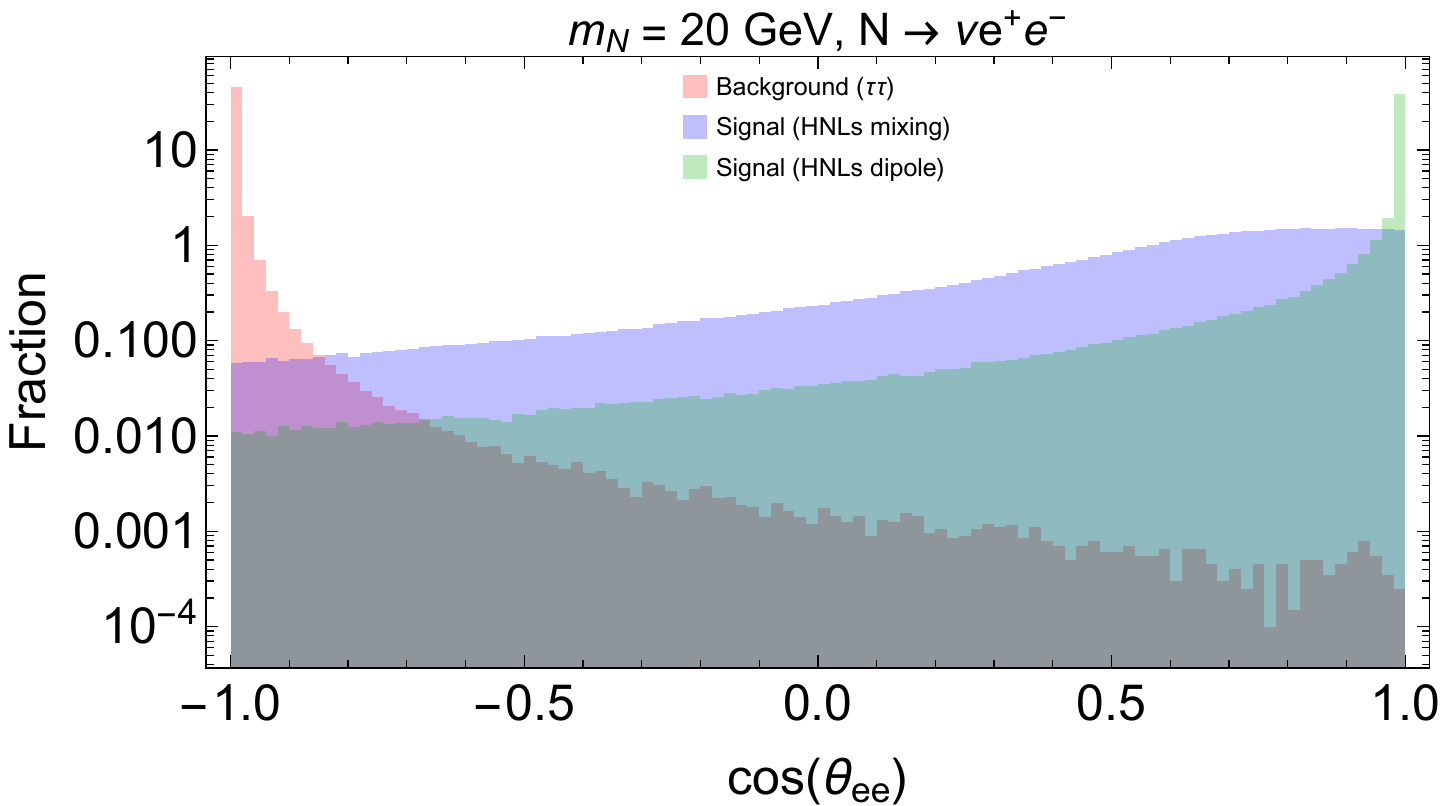}
    \caption{The distribution of the $e^{+}e^{-}$ pair in cosine of the angle between the $e^{+},e^{-}$ at FCC-ee in the $Z$-pole operating mode. Three processes are considered: the background process $Z\to \tau\bar{\tau}\to e^{+}e^{-}\bar{\nu}_{e}\nu_{e}\bar{\nu}_{\tau}\nu_{\tau}$, and the HNL decays $N\to e^{+}e^{-}\nu$, assuming the dipole and the mixing couplings (mixing with $\nu_{e}$ is considered), respectively. The detector reconstruction effects are not included.}
    \label{fig:cosine-distribution}
\end{figure}

The situation with the signal is different: the angle distribution between the $e^{+},e^{-}$ originated from the HNL decay is peaked at $\theta = 0$, and the situation remains the same even for heavy HNLs $m_{N}\simeq m_{Z}$. Therefore, the background yield may be reduced without a significant impact on the signal if one requires a cut on $\cos(\theta_{ee})$ and $E_{e^{+}},E_{e^{-}}$ from below.

To estimate the effect of such a cut on the background and signal, we have simulated $\simeq 5\cdot 10^{9}$ decays $Z\to \tau\bar{\tau} \to e^{+}e^{-}\nu_{e}\bar{\nu}_{e}\nu_{\tau}\bar{\nu}_{\tau}$, which corresponds to the full statistics expected during the full timeline of FCC-ee in the Z pole mode~\cite{Blondel:2022qqo}. In the simulation, we included neither finite detector reconstruction resolution\footnote{Nevertheless, as is demonstrated in~\cite{Blondel:2022qqo}, FCC-ee has perfect reconstruction capabilities of both the lepton energies and momentum (and hence $\cos(\theta_{ee})$).} nor the particle identification efficiency ideally. Therefore, its predictions should be validated with full-scale simulations. 

The distribution in $\cos(\theta_{ee})$ for the $e^{+}e^{-}$ pair from the background and the decays $N\to e^{+}e^{-}\nu$, considering both the models of the dipole portal and the minimal HNL model, is shown in Fig.~\ref{fig:cosine-distribution}. With the simulated sample, we have reproduced the selection efficiencies reported in Table 2 of~\cite{Blondel:2022qqo} for the process $Z\to \tau\tau\to ee$. Next, we found that the cut
\begin{equation}
    \cos(\theta_{ee}) > -0.5, \quad E_{e^{+}} > 2\text{ GeV}, \quad E_{e^{-}} > 2\text{ GeV}
    \label{eq:alternative-cuts-appendix}
\end{equation}
reduces the number of backgrounds to zero even before imposing the $d_{0}$ and $\slashed{p}$ cuts in the simplified simulation  mentioned above. 

The same conclusion may hold for other decays $Z\to f\bar{f}\to e^{+}e^{-}+\text{inv}$. The $e^{+},e^{-}$ pair originates either from the single process $f \to \dots \to e^{+}e^{-}+\text{inv}$\footnote{Here, if $f$ is a quark, by $\dots$ we mean a hadronization.} such that $\bar{f}\to \dots \to \text{inv}$ (the diagram (b) in Fig.~\ref{fig:events-topology}), or from the two independent processes $f\to \dots \to e^{-}+\text{inv}$, $\bar{f}\to \dots \to e^{+}+\text{inv}$ (the diagram (c)). By inspecting the decay modes of possible products of $f$ in~\cite{Workman:2022ynf} and assuming a perfect detector efficiency in detecting charged particles and neutral long-lived mesons such as $K^{0}_{L}$ (via deposition in HCAL), we have not found the combination $f\bar{f}$ which may lead to the diagram (b). Therefore, we conclude roughly that this category of events may be a subject of the detector inefficiency only. The level of this inefficiency is to be determined by the full-scale simulations. 

Nevertheless, we believe that the combination of the presented cuts in addition to the vertex criteria (such as the small distance of the closest approach between the tracks) would allow reducing the background from $Z$ boson SM decays to zero. Further, we will ideally assume zero background from the processes of the type $Z\to f\bar{f}\to Y\bar{Y}+\text{inv}$, constituting the background for various decay modes of the HNL $N\to Y\bar{Y}+\nu$. This may be especially the case for the decays e.g. $N\to q+\bar{q}+\nu$, for which the background process, $Z\to q+\bar{q}$, would have even more marginal kinematics.
Note that a more detailed simulation checking the selection rules we proposed in Eq.~\eqref{eq:alternative-cuts-appendix} is strongly suggested to be performed, which is beyond the reach of this work. 

\end{document}